\documentclass[showpacs,aps,pra,superscriptaddress,floatfix,preprint]{revtex4-1}
\usepackage[french,english]{babel} 
\usepackage[T1]{fontenc} 
\usepackage[utf8]{inputenc} 
\usepackage{lmodern}
\usepackage{microtype}
\usepackage{hyperref}
\usepackage{amsfonts, vmargin}
\usepackage{amsmath, amssymb,mathrsfs}
\usepackage{graphicx}
\usepackage{tikz}
\usepackage{epic, eepic}
\usepackage[us]{datetime}

\newcommand{\br}{{\bf r}}
\newcommand{\G}{{\bf \Gamma}}

\begin{document}

\title{``Phase diagram'' of a mean field game}

\author{Igor Swiecicki}
\affiliation{Laboratoire de Physique Th\'eorique et Mod\`eles
  Statistiques, CNRS UMR 8626, Univ. Paris-Sud,  91405 Orsay Cedex,
  France} 
\affiliation{Laboratoire de Physique Th\'eorique et Mod\'elisation,
  CNRS UMR 8089, Universit\'e   Cergy-Pontoise, 95011 Cergy-Pontoise
  Cedex, France}
\author{Thierry Gobron} 
\affiliation{Laboratoire de Physique Th\'eorique et Mod\'elisation,
  CNRS UMR 8089, Universit\'e   Cergy-Pontoise, 95011 Cergy-Pontoise
  Cedex, France}
\author{Denis Ullmo}
\affiliation{Laboratoire de Physique Th\'eorique et Mod\`eles
  Statistiques, CNRS UMR 8626, Univ. Paris-Sud,  91405 Orsay Cedex,
  France} 

\begin{abstract}

  {Mean field games were introduced by J-M.Lasry and P-L. Lions in the
    mathematical community, and independently by M. Huang and
    co-workers in the engineering community, to deal with optimization
    problems when the number of agents becomes very large. In this
    article we study in detail a particular example called the
    ``seminar problem'' introduced by O.Gu\'eant, J-M Lasry, and
    P-L. Lions in 2010.  This model contains the main ingredients of
    any mean field game but has the particular feature that all agent
    are coupled only through a simple random event (the seminar
    starting time) that they all contribute to form.  In the mean
    field limit, this event becomes deterministic and its value can be
    fixed through a self consistent procedure.  This allows for a
    rather thorough understanding of the solutions of the problem,
    through both exact results and a detailed analysis of various
    limiting regimes.  For a sensible class of initial configurations,
    distinct behaviors can be associated to different domains in the
    parameter space . For this reason, the ``seminar problem'' appears
    to be an interesting toy model on which both intuition and
    technical approaches can be tested as a preliminary study toward
    more complex mean field game models.  }
\end{abstract}

\date{\today}

\maketitle


\section{Introduction}

Many problems in different fields deal with a situation where many
identical and interacting agents try to minimize a cost through the
choice of a strategy. One can think of economic agents trying to
maximize their profits, of people in a crowd trying to minimize their
discomfort or to particles in a fluid ``trying'' to minimize their
energy.

A general framework making  possible to model a large class of such
problems has been introduced in 2006 by Lasry and Lions
\cite{LasryLions2006-1,LasryLions2006-2} and Huang et al.\
\cite{Huang2006} under the general terminology of ``mean
field game theory''.  Largely inspired by statistical physics, this
approach addresses the limit where the agents face a continuum of
choices (states) in which they can evolve only locally, and the number
of agent is large enough that self averaging processes are at
work. This approach leads to a system of partial differential equations
coupling the density of players and the optimization part of the
problem.

Mean field game theory has been intensively studied in the past few
years, and in spite of its relative youth, a very large number of
results have been obtained in the mathematical
\cite{Carmona-ctrl2013,DegondRinghofer2014,Cardaliaguet2013,AchdouCapuzzo2010,Cardaliaguet-course}
and socio-economic community
\cite{LachapelleWolfram2011,BesancDamCourtault2012,Carmona-MFG2013,Burger2013,BalandatTomlin2013}. A
recent overview is given by Gomes and Sa\'ude in
\cite{GomesSaude2014}.  Most of the focus however has been put either
on the conditions required to prove rigorously the existence and
unicity of the solutions of the equations of mean field game
theory \cite{Cardaliaguet-course}, or on the study of particular
models based primarily on numerical treatments \cite{AchdouCapuzzo2010}.  A more ``qualitative'' understanding of the behavior of
the solutions, based on the identification of the relevant time and
length scales, and on the analytical study of the solution in various
limiting regime, has received significantly less attention.

Our goal in this paper is to perform this program for a simple model,
introduced by Gu\'eant {\sl et al.} in 2010
\cite{GueantLasryLions2010}, called the seminar problem to be
described in more details below. The essential point here is that this
``mean field game model'' is in some sense very close to the everyday
``Physicists' mean field'' since all agents are interacting only
through a very simple ``field'' which is actually a simple number, the
time $T$ at which the seminar actually starts. This particular feature
allows for an analytical approach, similar in spirit to the
physicists' one: For fixed $T$, the behavior of each agent becomes
independent on the other, making the associated problem to be solvable
to a large extent; then, for a given distribution of agents, the
actual value of $T$ can be evaluated by a self-consistency procedure.
The main interest in this model is to provide a fully understandable
toy model on which one can develop its own intuition and tools before
tackling the full complexity of mean field game models.

The paper is organized as follows : In Section~\ref{sec:model} we
introduce the seminar problem in details and show that its resolution
involves two essentially independent parts : a system of coupled
(Hamilton-Jacobi-Bellman and Kolmogorov) differential equations on one
hand, and a self-consistency problem on the other.
Sections~\ref{sec:HJB} and \ref{sec:FP} address the
Hamilton-Jacobi-Bellman and Kolmogorov equations,
respectively. Various limiting regimes are studied in details for
both.  Moreover, we show that an exact solutions to these coupled
differential equations can actually be given in a closed form. The
self-consistency condition determining the effective beginning of the
seminar $T$ is discussed in Section~\ref{sec:SCR}, eventually leading
to the construction of a ``phase diagram'' for this toy model.
Concluding remarks are gathered in Section~\ref{sec:conclusion}. The
paper is completed by three Appendices where technical computations
are shown.
	

\section{The seminar problem}
\label{sec:model}

\subsection*{The model}

Consider a corridor at the end of which is a seminar room. A
  seminar is planned at time $\bar{t}$ but people know that in
  practice, it will only begin when a large enough proportion of the
  lab members $\theta$ (known), will be seated. 

 The members of the
  laboratory thus move according to the following considerations:
They do not want to arrive too early in the seminar room because they
do not particularly enjoy waiting idly as the room fills.  On the
other hand they are aware that the lab director and the seminar
organizers will already be in the room at time $\bar{t}$, and will
frown upon late comers.  Furthermore they really want to understand
the content of the seminar and are concerned that missing the actual
beginning might not help in this respect.

For every agent, this is summarized by the following cost function
associated with the arrival time $t$ :  
\begin{equation} \label{eq:c(t)}
 c(t) = \alpha [t-\bar{t} \ ]_+ + \beta [t-T]_+ + \gamma [T-t]_+ \; ,
 \end{equation}
 where $T$ is the effective beginning time of the seminar.  In
 Eq.~(\ref{eq:c(t)}), $\alpha$, $\beta$ and $\gamma$ are positive real
 numbers and respectively quantify the sensitivity to social pressure,
 the desire not to miss the beginning of the seminar, and the
 reluctance to useless waiting.  We assume these parameters to be the
 same for all members of the laboratory.  We also assume $(\gamma <
 \alpha)$ so that the cost $c(t)$ is actually minimal for the official
 starting time $\bar t$.

 The corridor is represented by the negative half-line line
   $\mathbb{]} -\infty, 0 \mathbb{]}$, and the seminar room is located
   at $x=0$. At time $t=0$, people leave their office to go to the
   seminar. Each member of the laboratory $i=1 \dots N$, controls her
 drift $a_i(t)$ toward the seminar room but is subject to random
 perturbation (stopping to discuss with somebody, going back to
   take a pen and then giving up the idea, or speeding up to catch up
   a friend for example), modelled by a Gaussian white noise of
 variance $\sigma^2$. A given participant thus moves according to a
 noisy dynamics:
 \begin{equation} \label{eq:langevin}
\mathrm{d} X_i = a_i(t) \, \mathrm{d}t +\sigma \mathrm{d} W_i(t)
 \end{equation}
where,
 \begin{align*}
	& X_i(t) \mbox{ is the  agent position at time $t$ } \; , \\
	& a_i(t) \mbox{ is her  drift at the same time} \; , \\
	& \mathrm{d} W_i(t) \mbox{ is a normal white noise} \; .
 \end{align*} 
 Again, except for their initial positions, all
   agents have the same characteristics.

   In addition to the cost $c(t)$ associated to the arrival time
   (Eq.~(\ref{eq:c(t)})), agents dislike having to rush on their way
 to the seminar room and the total cost function therefore includes a
 terms quadratic in the (controlled) drift $a_i(t)$.  An agent leaving
 her workplace $x_0$ at $t=0$ has thus to adapt her drift in order to
minimize the expected cost
\begin{equation} \label{eq:total_cost}
	J_T[a]=\mathbb{E} \left [ c(\tilde{\tau}) + \frac{1}{2}
          \int_0^{\tilde{\tau}} a^2_i(\tau) \ \mathrm{d}\tau \right ]
        \, 
\end{equation}
associated with Eq.~\eqref{eq:langevin} and the initial
  condition $X(t \!= \! 0) = x_0 \!<\!0$. In Eq.~(\ref{eq:total_cost})
  $\tilde{\tau}$ is the first passage time at $x=0$
\begin{equation}
\tilde{\tau}= \inf\{ t: X(t) =0\} \; ,
\end{equation}
and $\mathbb{E}$ is the expectation  with respect to the  noise.

We define $N(t)$ as the cumulative distribution of arriving time (the
percentage of people arrived before $t$).  If the quorum is met before
the official time $\bar t$ of the seminar, this latter starts exaclty
at $\bar t$.  If on the other hand the quorum is met at a later time,
$T$ is determined by the self-consistency relation $N(T)=\theta$ (more
formally $T = \inf\{t\ge \bar t : N(t)\ge
\theta\}$).\footnote{Strictly speaking, the effective starting time
  $T$ is a random event (a random stopping time). Here we assume that
  in the mean field limit of the present model, this event becomes
  deterministic so that we can confuse it with its expectation for
  almost all realizations.}

Within the mean field approximation, the total number of researchers
in the lab is assumed to be large enough that the
 individual choices of a 
 given agent, and thus her arrival time, cannot have any significant
impact on $T$.  Each agent should thus solve the optimization problem
Eqs.~(\ref{eq:langevin})-(\ref{eq:total_cost}) for herself, assuming $T$ fixed.
Introducing the value function
\begin{equation} \label{eq:value_function}
	u(x,t) = \min_{a_i(t)} \left\{ \mathbb{E} \left [
              c(\tilde{\tau}) + \frac{1}{2} 
          \int_t^{\tilde{\tau}} a^2_i(\tau) \ \mathrm{d}\tau \right ] \right\}
\end{equation}
subject to the initial condition $X(t) = x$, this optimization problem
is equivalent to the Hamilton-Jacobi-Bellman (HJB) equation (see
  e.g. \cite{KampenBook}):
\begin{equation} \label{eq:HJB}
\left\{
\begin{aligned}
& \frac{\partial u}{\partial t} -\frac{1}{2} \left( \frac{\partial
  u}{\partial x} \right)^2+\frac{\sigma^2}{2} \frac{\partial^2 u}{\partial
  x^2}=0 \\ 
& u(x=0,t) =c(t)
\end{aligned}
\right. \; ,
\end{equation}
and the optimal drift is given by 
\begin{equation} \label{eq:drift}
a(x,t) = - \partial_x u(x,t) \; .
\end{equation}

The second hypothesis underlying the mean field approximation is that,
beyond the total size of the agent population, the agent
  density itself is large enough that the distribution of agents is
self-averaging:  therefore  everything happens as if at any given
location and time, each realization of the noise was experienced by
somebody.  Assuming a (normalized) initial density of participants
$m_0(.)$ at time $t \!=\!  0$, Eq.~(\ref{eq:langevin}) thus implies
that this density will evolve under the Kolmogorov equation (see
  e.g. \cite{Fleming&SonerBook}):
\begin{equation} \label{eq:KMG}
\left\{
\begin{aligned}
 \frac{\partial m}{\partial t}+\frac{\partial am}{\partial
   x}-\frac{\sigma^2}{2} \frac{\partial^2 m}{\partial x^2}=0 \\ 
 m(x=0,t)=0 \\
 m(x,t\!=\!0) = m_0(x)
\end{aligned}
\right. \; .
\end{equation} 
Once this equation is solved, the quorum condition 
\begin{equation} \label{eq:SCC}
N(T) = \left[ 1 - \int_{-\infty}^0 m(x,T)\right]  
 = \theta   \qquad \mbox{    (if $T > \bar t$)} 
\end{equation} 
(or $N(T)\geq \theta$ if $T = \bar t$) 
provides a self-consistent condition which has to be fulfilled if $T$
is indeed the actual starting time of the seminar.

\subsection*{General strategy}

The toy model that we just described depends on a few parameters which
play different roles.  Some of these parameter are just ``numbers''.
For instance the official time of the seminar $\bar{t}$, which mainly
fixes a time scale.  Or the parameters $\alpha$, $\beta$ and $\gamma$
of the cost function $c(t)$ Eq.~(\ref{eq:c(t)}) which, as we shall
see, govern the typical amplitude of the drift velocity.  In the same
way the noise strength $\sigma$ will govern the diffusion velocity.

Another parameter of the problem is the initial distribution of agents
$m_0(x)$. Being a function rather than just a number it is a little
bit more difficult to characterize simply.  It defines a mean initial
position $\langle x \rangle_0$, but also moments of arbitrary order,
which may introduce various length scales into the problem.

We are helped here by the linear character of the Kolmogorov
equation. Indeed introducing the elementary solutions $G(x,t|x_0)$ which
are the solutions of Eq.~(\ref{eq:KMG}) with a Dirac mass
$\delta(x-x_0)$ as initial condition, the solution for an arbitrary
$m_0(x)$ is obtained through the convolution 
\[ 
m(x,t) = \int_{-\infty}^0 dx_0 \, G(x,t | x_0) m_0(x_0) \; .
\]
Therefore, introducing 
\begin{equation} \label{eq:rho}
\rho(x_0,t) \equiv \int_{-\infty}^0 dx \,G(x,t | x_0) \; ,
\end{equation}
which thus measure the proportion of agents starting from $x_0$ who
have not yet reached the seminar room at time $t$, the self-consistent condition
Eq.~(\ref{eq:SCC}) reads
\begin{equation} \label{eq:SCC2}
\int_{-\infty}^0dx_0 \, \rho(x_0,T) m_0(x_0) = \bar \theta \qquad
\mbox{(if $T > \bar t$)}\; , 
\end{equation}
with $\bar \theta  = (1-\theta)$ the proportion of agents still in
the corridor when the quorum is met.

The solution of the self consistent problem can therefore be split
quite neatly in two distinct parts.  The first part will be to
analyze, and solve, the Hamilton-Jacobi-Bellman and Kolmogorov
equations (\ref{eq:HJB}) and (\ref{eq:KMG}) assuming $T$ known.  More
specifically, the goal in this first part will be to compute the
function $\rho(x_0,T)$ for arbitrary $x_0$ and $T$.  This is what we
shall do in the two following sections.  For this part we obviously do
not need to specify what is $m_0(x_0)$.

Once $\rho(x_0,T)$  is known, the  self consistent problem  reduces to
Eq.~(\ref{eq:SCC2}).  It  then of  course involves the  initial density
$m_0(x_0)$,  as  well  as  $\rho(x_0,T)$,  but  this  latter  quantity
summaries all  the   required information, beyond $m_0(x_0)$,  about the
system.  This second aspect of the problem will be addressed in
  section~\ref{sec:SCR}.

\section{Resolution of the Hamilton-Jacobi-Bellman equation}
\label{sec:HJB}

Except for its rather non-standard boundary conditions, the HJB
  equation Eq.~(\ref{eq:HJB}) is closely related to a Burger's equation, and
  it can in the same way be solved exactly through a rather standard
  Cole-Hopf transformation. Before we do so however, we find it useful
  to consider first the limiting behaviors of very small and very
  large $\sigma$'s.

\subsection{Small \texorpdfstring{$\sigma$}{sigma}}

To understand the regime of very weak noise, let us consider the
noiseless limit of  Eq.~(\ref{eq:HJB}), which takes the form of
the  Hamilton-Jacobi equation
\begin{equation} \label{eq:HJ}
L( \partial_t u, \partial_x u) = 0 \; ,
\end{equation}
associated with the free propagation Hamiltonian
\begin{equation} \label{eq:Hamiltonian}
L(E,p) \equiv E -\frac{p^2}{2} \,
\end{equation}
complemented with the  boundary conditions
\begin{equation} \label{eq:HJ-bc} 
 u(x\!=\!0,t) =c(t) \; .
\end{equation}

Introducing a fictitious time $\xi$ and noting $\dot{(\;)} =
d{(\;)}/d\xi$ the corresponding time derivative, the Hamilton dynamics
associated with $L$ is given by \cite{ZaudererBook,MaslovBook}
\begin{equation}\label{eq:Hamilton}
\begin{aligned}
\dot t = \frac{\partial L}{\partial E} = 1 \; &
\qquad \dot E = - \frac{\partial L}{\partial t} = 0 
\\
\dot x= \frac{\partial L}{\partial p} = -p \; &
\qquad \dot p = - \frac{\partial L}{\partial x} = 0 \; .
\end{aligned}
\end{equation}

Solution of the  Hamilton-Jacobi equation are typically obtained
through me method of characteristics.  Here, this amounts to build a
one parameter family of rays $\br_{\tilde \tau}(\xi)$ ($ \br\equiv
(E,t,p,x)$) indexed by $\tilde \tau$, such that $\br_{\tilde
  \tau}(\xi)$ is solution of the Hamilton's equations
Eq.~(\ref{eq:Hamilton}), and with initial conditions  $\br(\xi\! =\!
0) = (E_0,t_0,p_0,x_0)$ imposed by  Eq.~(\ref{eq:HJ-bc}) as
\begin{eqnarray*}
t_0 & = & \tilde \tau \\
x_0 & = & 0 \\
E_0 & = & \frac{dc}{dt}(\tilde \tau) \equiv c'(\tilde \tau)\\
L(E_0,p_0) & = & 0 \; .
\end{eqnarray*}
This last equation imposes
\begin{equation} \label{eq:v0} 
p_0 = - \sqrt{2 c'(\tilde \tau)} \; .
\end{equation}
Once this one parameter family of ray is build,  the solution of the
Hamilton-Jacobi equation just reads (for $t < \tilde \tau$, and thus
negative $\xi$)
\begin{eqnarray}
u(x\!=\!x_{\tilde \tau}(\xi),t\!=\!t_{\tilde \tau}(\xi)) & = & c(\tilde \tau) +
\int_0^\xi (E \dot t + p \dot x) d\xi \nonumber \\
& = &c(\tilde \tau) - (\tilde \tau - t) c'(\tilde \tau)  - 
x \sqrt{2 c'(\tilde \tau)}  \; . \label{eq:action}
\end{eqnarray}

\begin{figure}[ht]
\includegraphics[width=10cm]{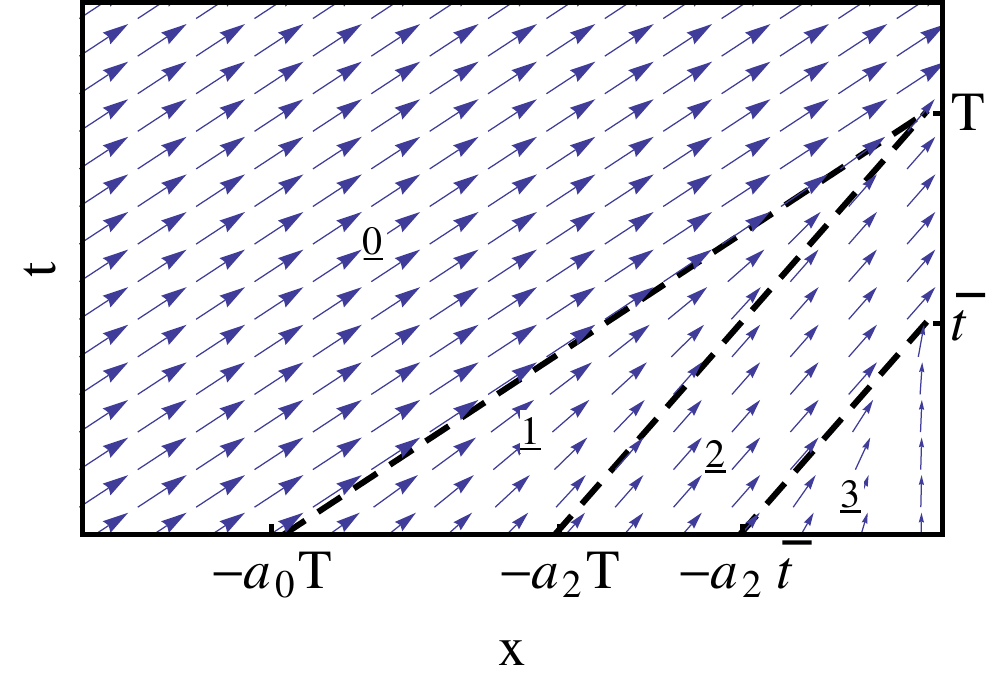}
\caption{Regions of the space $(x,t)$ ($x<0, t>0$)
and the corresponding  drift fields for $\sigma \to 0$ (cf
Eq.~(\ref{eq:a_small_sigma})) [$a_0=\sqrt{2
    (\alpha+\beta)}, a_2=\sqrt{2( \alpha-\gamma )}$].  
 }
\label{fig:HJ}
\end{figure}

As illustrated on Fig.~\ref{fig:HJ} the quarter plan  ($x<0, t>0$) has
to be divided in four different regions, 
\begin{equation} \label{eq:regions}
\begin{cases}
	  \mbox{Region (0) :   } &  x \leqslant -\sqrt{2(\alpha +\beta)}(T-t) \\
	  \mbox{Region (1) :    }& -\sqrt{2(\alpha
            +\beta)}(T-t) \leqslant x \leqslant
          -\sqrt{2(\alpha-\gamma)}(T-t) \\ 
	 \mbox{Region (2) :   } & \sqrt{2(\alpha-\gamma)}(T-t)
         \leqslant x \leqslant -\sqrt{2(\alpha-\gamma)}(\bar{t}-t)
         \\ 
          \mbox{Region (3) :   }  &
          -\sqrt{2(\alpha-\gamma)}(\bar{t}-t) \leqslant x <0 \; ,
\end{cases} 
\end{equation}
for which the application of Eq.~(\ref{eq:action}) is somewhat
different.  In region (0) for instance the relevant rays reach $x=0$
at $\tilde \tau > T$ which corresponds to $c'(\tilde \tau) =
(\alpha+\beta) \equiv c'_0$. In the same way for region (2) $\bar t <
\tilde \tau < T$ and $c'(\tilde \tau) = (\alpha - \gamma) \equiv
c'_2$.

Region (1) then corresponds to $\tilde \tau = T$, where $c'(\tilde
\tau)$ is discontinuous.  It can be easily justified (e.g.\ by viewing
$c(t)$ as the limit of a family of differentiable functions) that the
correct procedure here is to use all the rays emerging from
$(x\!=\!0,t\!=\! T)$ with all possible values of $c'(\tilde \tau)$ within
the interval $] (\alpha - \gamma),(\alpha+\beta)[ $ (and thus all
velocities $p_0$  within $] \sqrt{2(\alpha -
  \gamma)},\sqrt{2(\alpha+\beta)}[)$. 

In the same way region (2)  corresponds to $\tilde \tau = \bar t$, and
one should use  all the rays emerging from
$(x\!=\!0,t\!=\! \bar t)$ with all possible values of $c'(\tilde
\tau)$ within $] 0, (\alpha - \gamma)[ $ (and thus all
velocities $p_0$ within  $]0,\sqrt{2(\alpha -  \gamma)}[)$. 

Note that because $c'(\tilde \tau)$ is negative for $\tilde \tau <
\bar t$, Eq.~(\ref{eq:v0}) has not real solution for $p_0$ and it is
not possible to fulfil the boundary conditions Eq.~(\ref{eq:HJ-bc}) in
this time interval for the Hamilton-Jacobi equation.  In a small layer
near the line, $(x=0,0<t<\bar t)$, the fact the Hamilton-Jacobi
equation is first order when the HJB equation is second order implies a
qualitative difference between the limit of small $\sigma$'s and
$\sigma=0$.

Acknowledging this, and keeping in mind the procedure explained above to
handle the discontinuities of $c'(\tilde \tau)$, Eq.~(\ref{eq:action})
gives an explicit solution for the Hamilton-Jacobi equation
Eq.~(\ref{eq:HJ})-(\ref{eq:HJ-bc}).  What will be needed though as an
input for the Kolmogorov equation Eq.~(\ref{eq:KMG}) is not so much
$u(x,t)$ itself than its spatial derivative $-\partial_x u$ which
through Eq.~(\ref{eq:drift}) specifies the drift $a(x,t)$ in
Eq.~(\ref{eq:KMG}).   From Eq.~(\ref{eq:action}) we see that
$-\partial_x u$ is just the velocity $-p_0$ of the free motion on the
corresponding ray.  We obtain therefore the following results,
\begin{equation} \label{eq:a_small_sigma}
 a(x,t) = - \partial_x u(x,t)=
\begin{cases}
	  \sqrt{2(\alpha+\beta)} \equiv a_0  & \mbox{ in Region (0) } \\
	  \displaystyle \frac{-x}{(T-t)} & \mbox{ in Region (1) } \\
	 \sqrt{2(\alpha-\gamma)}  \equiv a_2 & \mbox{ in Region (2) } \\
  \displaystyle \frac{-x}{(\bar{t}-t)} & \mbox{ in Region (3) } \\
  \end{cases}  
\end{equation}
 valid as  ${\sigma \to 0}$. On Fig.~\ref{fig:HJ}, this
   velocity is shown  as the inverse slope of the arrows.   

In the deterministic limit considered in this subsection a
  finite fraction of the agents (namely all those starting in region
  (1)) arrive exactly at time $T$.  The quorum condition
  Eq.~(\ref{eq:SCC}) is therefore ill-defined in this limit.

\subsection{Large \texorpdfstring{$\sigma$}{sigma}}

Let us consider now the HJB equation Eq.~(\ref{eq:HJB}) in the limit
of very large $\sigma$'s.  This amounts here to neglect the nonlinear
term $\frac{1}{2} (\partial_x u)^2$, yielding the backward diffusion
equation  
\begin{equation} \label{eq:HJB-large-sigma}
\left\{
\begin{aligned}
& \frac{\partial u}{\partial t} + \frac{\sigma^2}{2} \frac{\partial^2
  u}{\partial   x^2}=0 \\ 
& u(x=0,t) =c(t) \; .
\end{aligned}
\right.
\end{equation}
In this subsection it will be convenient to use for the
boundary conditions a slightly modified version $c_\Lambda(t)$ of the
cost function Eq.~(\ref{eq:c(t)}),
\begin{equation} \label{eq:tilde_c}
\left\{
\begin{aligned}
&c_\Lambda(t) =& c(t)    \qquad &\mbox{for } t \leq \Lambda \\ 
&c_\Lambda(t) =& c(\Lambda) \qquad &\mbox{for } t \geq \Lambda \\ 
\end{aligned}
\right.
\end{equation}
with ($\Lambda \gg \bar t, T$) a very large time (one may imagine for
instance that once the seminar is over, there is less marginal
incentive to reach the seminar room).

There are many ways to derive a solution of Eq.~(\ref{eq:HJB}), but a
relatively transparent one consists in going back to the original
optimization problem, i.e. to define $u(x,t)$ as
Eq.~(\ref{eq:value_function}).  Indeed, in the limit of very large
$\sigma$'s, this optimization is straightforward: if the motion is
overwhelmingly dominated by the noise, the best strategy for an agent is
just to renounce paying the cost of the drift, and hope that the
diffusive motion will bring her in time for the seminar.

Let us note
\begin{equation} \label{eq:G0}
G_0(x,t) \equiv \frac{1}{\sqrt{2 \pi \sigma^2 t}} \exp
\left(-\frac{x^2}{2\sigma^2 t} \right)
\end{equation}
the elementary solution of the free diffusion problem. The
distribution of time of first passage at $x=0$ for free diffusion
started at $t_0$ in $x_0<0$ is given by \cite{KampenBook}
\[
P(t) = - \frac{d}{dt} \int_{+x_0}^{-x_0} dx \, G_0(x,t-t_0)
\]
The value function $u(x,t)$ is just the average of the cost
  function $\tilde c(t) = c_\Lambda(t)$ with  this first
passage time distribution. It therefore reads
\begin{align} 
\lim_{\sigma \to \infty}u(x_0,t_0) & =   \int_{t_0}^\infty dt \, \tilde c(t) P(t)
\label{eq:u-large-sigma:a}\\
& =  - x_0 \int_0^\infty dt \, \frac{\tilde c(t+t_0)}{t}
G_0(x_0,t) \; , \label{eq:u-large-sigma}
\end{align}
 [The fact that $G_0(x,t)$ is the
elementary solution of the diffusion equation has been used to
transform (\ref{eq:u-large-sigma:a}) into (\ref{eq:u-large-sigma})].

Note that Eq.~(\ref{eq:u-large-sigma}) would be valid for any choice
of the final cost function $\tilde c(t)$ as long as the integral
converges in $+\infty$, i.e. as long as $\tilde c(t)$ grows less than
linearly at infinity.  Thus the need to modify the large $t$ behavior
of $c(t)$ in this subsection.

\subsection{Arbitrary \texorpdfstring{$\sigma$}{sigma}}

As was mentioned at the beginning of this section, the HJB equation
can actually be solved for arbitrary values of $\sigma$.  Indeed,
using the Cole-Hopf transformation i.e. setting $u(x,t)= - \sigma^2 \ln
\phi (x,t)$ yields a linear equation for $\phi (x,t)$:
\begin{equation}
\left\{
\begin{aligned}
& \frac{\partial \phi}{\partial t} +\frac{\sigma^2}{2}
\frac{\partial^2 \phi}{\partial x^2}=0 \\ 
& \phi(x=0,t)=e^{-\frac{c(t)}{\sigma^2}}
\end{aligned}
\right.
\end{equation}
It's solution is thus the same as Eq.~(\ref{eq:u-large-sigma}) with
$\tilde c(t) \equiv \exp{[-{c(t)}/{\sigma^2}]}$ as cost function. 
We obtain in this way
\begin{align}
\phi(x,t)=&   -x \int_0^{\infty}
 \frac{e^{-\frac{c(t+\tau)}{\sigma^2}}}{\tau}G_0(x,\tau) \
\mathrm{d}\tau   \label{eq:phi-general} \\ 
u(x,t)=&-\sigma^2 \ln \phi(x,t)  \; . \label{eq:u-general} 
\end{align}
An explicit expression of $\phi(x,t)$ in terms of elementary functions
is given in Appendix~\ref{app:ExSol} (see Eq.~(\ref{eq:phi-full})).
We just stress here that, because for $t$ larger than $T$ the cost
function Eq.~(\ref{eq:c(t)}) becomes linear ($c(t) = (\alpha+\beta)t
-(\alpha \bar t + \beta T)$), $\phi(x,t)$ takes a particularly simple
form (see Eq.~(\ref{eq:phi-simple})) from which the value function is
deduced as
\begin{equation} \label{eq:u-simple}
u (x,t\!>\!T) =  - \sqrt{2 (\alpha +\beta)} x - c(t)  \; .
\end{equation}
As a consequence, for times beyond $T$ the drift $a(x,t)$ is just the
constant 
\begin{equation} \label{eq:a-simple}
a (x,t\!>\!T) =  \sqrt{2 (\alpha +\beta)} = a_0 \; .
\end{equation}

From Eqs.~(\ref{eq:phi-general})-(\ref{eq:u-general}), the
  limiting behaviors
  Eqs.~(\ref{eq:a_small_sigma})-(\ref{eq:u-large-sigma}) can be
  recovered. This is particularly simple, for instance,   in the large
$\sigma$ regime if one uses the regularized version $c_\Lambda(t)$ of
  the cost function.  Indeed in that cas  we see that as soon as $\sigma^2 \gg
c(\Lambda)$, we can expand the exponential in
Eqs.~(\ref{eq:phi-general}) and the logarithm in
Eqs.~(\ref{eq:u-general}), and $u(x,t)$ reduces to
(\ref{eq:u-large-sigma}) in lowest order in $1/\sigma^2$.
Things are slightly trickier for the true (non regularized) cost
function $c(t)$ since however large $\sigma$ maybe, the $+\infty$
limit of the integral in Eqs.~(\ref{eq:phi-general}) is such that
$c(t+t_0) \gg \sigma^2$ (which actually simply provides an effective
cutoff for the integral).  We find in this case (see
appendix~\ref{app:phi-largesigma}) that   
\begin{equation} \label{eq:phi-largesigma}
	\phi(x,t)=\exp\left[-\frac{1}{\sigma^2} |x|\sqrt{2
              (\alpha+\beta)}{- c(t)}  \right] + O(\sigma^{-2})\; ,
\end{equation}
implying $a(x,t) = \sqrt{2 ( \alpha+ \beta)}$.
(If one consider furthermore the diffusive regime $a_0|x| \gg
\sigma^2$, where the above approximation is most useful, one can
further show that Eq.~(\ref{eq:phi-largesigma}) is valid up to
$O(\sigma^{-3})$ corrections.) 

When $\sigma^2 \to 0$, the integral in Eq.~(\ref{eq:phi-general}) can
be approximated using the steepest descent approximation in regions
(0) and (2); or noting that it is dominated by the boundary
contributions at $ t+t_0=T$ in regions (2) or $t+t_0=\bar t$ in region
(3) (see Eq.~(\ref{eq:regions}) or Fig.~\ref{fig:HJ}). Details of the
calculations and the precise condition under which the approximation
applies are given in appendix~\ref{app:phi-smallsigma}. The drif
velocity which can be expressed as $a(x,t) = \sigma^2 \partial_x\phi
/\phi$ is computed along the same lines, and  one recovers
in this way exactly Eq.~(\ref{eq:a_small_sigma}), except for a small
region near $\{ x=0 ; 0 \leqslant t \leqslant
\bar t \} $  scaling as $\sigma^2$ in region (0) and (2) and as
$\sigma$ in region (1) and (3).

\begin{figure}[ht]
\includegraphics[width=7 cm]{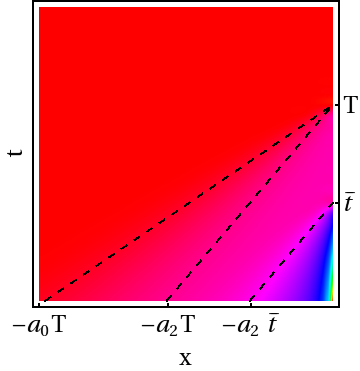}\\
\includegraphics[width=7 cm]{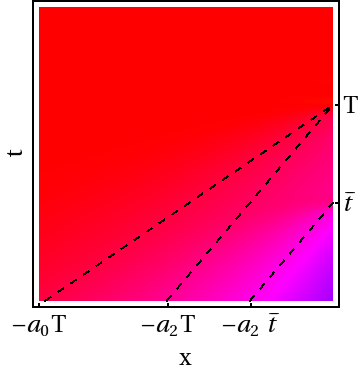}\\
\includegraphics[width=7 cm]{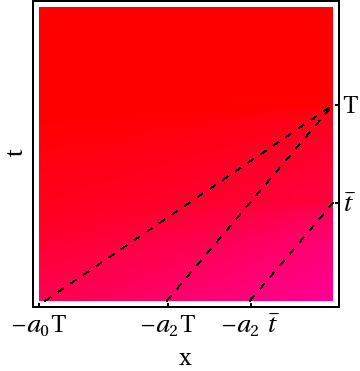}\\
\caption{ Evolution of the drift with $\sigma$ in heat
  representation. Top:$\sigma=0.5$, middle: $\sigma=2$, bottom:
  $\sigma=7$  [$\alpha=2, \beta=1, \gamma=1, \bar{t}=1, T=2$]} 
\label{fig:drift}
\end{figure}

To conclude this section, we note that the drift $a(x,t)$ obtained
from the agents' optimisation is exactly $\sqrt{2 ( \alpha+ \beta)}$
for $t > T$, but is actually close to this value in the entire
region~(0) already for small $\sigma$'s.  As $\sigma$' increases, the
part of the domain for which $a(x,t) \simeq \sqrt{2 ( \alpha+
  \beta)}$ increases beyond region~(0), and extends to essentially all
positions and times for large $\sigma$'s.  This evolution of the drift
with $\sigma$ is illustrated in Fig.~\ref{fig:drift}.

\section{Resolution of the Kolmogorov equation}
\label{sec:FP}

We turn now to the resolution of the Kolmogorov equation.  As we shall
see below, this equation can, in the particular case we consider here,
also be solved exactly.  Before proceeding to the description of this
exact solution, we find it useful nevertheless once again to discuss
briefly the two limiting cases.

\subsection{Limiting cases}
\label{subsec:lim}

\subsubsection{Case \texorpdfstring{$\sigma^2 \to 0$}{sigma}}

If we just set $\sigma$ to $ 0$, the Kolgomorov equation reduces to: 
\begin{equation} \label{eq:KMG0}
\left\{
\begin{aligned}
& \frac{\partial m}{\partial t}+\frac{\partial (a(x,t) m)}{\partial x}=0
\\ 
& m(x,t=0)=m_0(x)
\end{aligned}
\right. \; ,
\end{equation}
with the velocity $a(x,t)$ given by Eq.~(\ref{eq:a_small_sigma}).

Noting $\frac{D}{Dt}$ the total derivative attached to the flow
$a(x,t)$, Eq.~(\ref{eq:KMG0}) reads $Dm/Dt = 0$ in regions (0) and (2)
of Fig.~\ref{fig:HJ}, $Dm/Dt = -x/(T-t)$ in region (1), and  $Dm/Dt =
-x/(\bar t-t)$ in region (3).  This yields (see e.g. \cite{ZaudererBook})
\begin{equation} \label{eq:m0_small_sigma}
m(x,t) =
\begin{cases}
	  m_0 \left(x - \sqrt{2(\alpha+\beta)}t \right) & \mbox{ in
            Region (0) } \\ 
	  \frac{T}{T-t} m_0\left(x\frac{T}{T-t} \right) & \mbox{ in
            Region (1) } \\ 
	 m_0 \left(x - \sqrt{2(\alpha - \gamma)} t \right)& \mbox{ in
           Region (2) } \\ 
           \frac{\bar t }{\bar t -t} m_0\left(x\frac{\bar t }{\bar t
               -t}  \right)  & \mbox{ in Region (3) } 
  \end{cases}  \qquad .
\end{equation}
In other words, all agents starting at $t=0$ from a position $x_0 < -
\sqrt{2(\alpha+\beta)} T$ will arrive after $T$, all agents starting at
  $t=0$ from a position $x_0 > -\sqrt{2(\alpha-\gamma)} T$ will arrive
    before $T$, and all agents between $- \sqrt{2(\alpha+\beta)} T$ and
      $ -\sqrt{2(\alpha-\gamma)} T$ will arrive {\em exactly at
          time} $T$.  Therefore, at $\sigma=0$, the function
        $\rho(x_0,t)$ needed to define the self consistent condition
        Eq.~(\ref{eq:SCC2})
        become singular at $t=T$.  We shall see below how this
        behavior is regularized for a small but finite $\sigma$.

\subsubsection{Case \texorpdfstring{$\sigma^2 \to +\infty$}{sigma}}

When $\sigma^2 \to +\infty$, we have seen that the drift velocity
$a(x,t)$ tends toward the constant value $a_0 =
\sqrt{2(\alpha+\beta)}$. Intuitively, this is due to the fact that
when optimizing her drift, the agent does so not so much having in
minds the median arriving time but rather try to ensure herself
against possible late arrival due to the noise.  In this limit the
motion of the agents become extremely simple, as the mass of
participants is transported by an advection-diffusion equation with
constant drift $a_0$ and diffusion coefficient
$\frac{\sigma^2}{2}$. Forgetting for now the small technicalities
existing for small $|x|$, this would imply that we should simply
consider the Kolmogorov equation with constant drift
\begin{equation} \label{eq:KMG-large}
\left\{
\begin{aligned}
 \frac{\partial m}{\partial t}+ a_0 \frac{\partial m}{\partial
   x}-\frac{\sigma^2}{2} \frac{\partial^2 m}{\partial x^2}=0 \\ 
 m(x=0,t)=0 \\
 m(x,t=0)=m_0(x)
\end{aligned}
\right. \qquad .
\end{equation}
 Explicit solutions of Eq.~(\ref{eq:KMG-large}) are well
known (\cite{RednerBook}), and in particular the elementary solution for an
initial distribution $m_0(x) = \delta(x-x_0)$ is given by
\begin{equation}
G^{\rm CD}(x,t|x_0)=\frac{1}{\sigma \sqrt{2 \pi t}} \left\{ \exp \left(
    {-\frac{(x-x_0-  a_0 t)^2}{2 \sigma^2 t}} \right) 
    - \exp \left( {\frac{2a_0x}{\sigma^2}} \right)
       \exp \left( {\frac{(x+x_0+ a_0 t)^2}{2
      \sigma^2 t}} \right)
                          \right\} \; .
\label{eq:Green-CD}
\end{equation}
We shall see below that this expression indeed provides the leading
large $\sigma$ asymptotic approximation of the true 
solution.

\subsection{Full resolution of the coupled problem}

We turn now to the solution of the Kolmogorov equation for an arbitrary
$\sigma$.  For this purpose, let us write the agent density as
\cite{gueant2011} 
\begin{equation} \label{eq:IMT}
 m(x,t) = e^{-u(x,t) /\sigma^2} \G(x,t) \; ,
\end{equation}
with $u(x,t)$ the solution of the (HJB) equation Eq.~(\ref{eq:HJB}),
which is thus such that $- \partial_x u = a(x,t)$.  Inserting
Eq.~(\ref{eq:IMT}) into  Eq.~(\ref{eq:KMG}), we find that: 
\begin{equation}\label{eq:KMG-mod1}
\sigma^2 \partial_t \G - \frac{\sigma^4}{2} \partial_{xx}^2 \G = 
 \G \left(
  \frac{\partial u}{\partial t} -\frac{1}{2} \left(\frac{\partial
      u}{\partial x} \right)^2+\frac{\sigma^2}{2} \frac{\partial^2
    u}{\partial x^2} \right). 
\end{equation}
But $u(x,t)$ is a solution of the (HJB) equation Eq.~(\ref{eq:HJB}).
The right hand side of Eq.~(\ref{eq:KMG-mod1}) is therefore uniformly
zero, and this equation can be written as a simple diffusion equation
without drift
 \[
\left\{
 \begin{aligned}
&\frac{\partial \G}{\partial t} -\frac{\sigma^2}{2} \frac{\partial^2
  \G}{\partial x^2}=0 \\ 
& \G(x\!=\!0,t)=0 \mbox{ and }
\G(x,t\!=\!0)=e^{\frac{u(x,t\!=\!0)}{\sigma^2}}m_0(x) 
\end{aligned}
\right. \; .
 \]
Noting
 \begin{equation}\label{eq:G-abs}
 G^{\rm abs} _0(x,t|x_0) = \left( G_0(x,t|x_0)-G_0(x,t|-x_0)
 \right)  
\end{equation}
the elementary solution (Green's function) for the diffusion without
drift but with absorbing boundary in zero, obtained  straightforwardly
using the  method of image from the elementary solution of the free diffusion
problem Eq.~(\ref{eq:G0}), we find
\[
\G(x,t)=\int_{-\infty}^0 G^{\rm abs} _0(x,t|x_0) e^{u(x_0,t\!=\!0) /
  \sigma^2} m_0(x_0) \, \mathrm{d}x_0 \; .
\]
Inserting the expression Eq.~(\ref{eq:u-general}) of  $u(x,t)$ yields
\begin{equation} \label{eq:m-final}
m(x,t)=  \phi(x,t) \displaystyle
\int_{-\infty}^0 \frac{G_0^{\rm abs}(x,t|x_0)}{\phi(x_0,t\!=\!0)} m_0(x_0) \
\mathrm{d}x_0 \; ,
\end{equation}
and in particular, setting the initial distribution as a Dirac mass
located in $x_0$, we get for the elementary solution
\begin{equation} \label{eq:G-final}
G(x,t|x_0)=\frac{\phi(x,t)}{\phi(x_0,t\!=\!0)} \times G_0^{\rm abs}(x,t|x_0) \; .
\end{equation}

The self consistence equation Eq.~(\ref{eq:SCC2}) only involves
$\rho(x_0,t \!=\! T)$, and thus we  need to compute $G(x,t|x_0)$
at $t=T$, which is in the range for which $\phi$ can be expressed
through Eq.~(\ref{eq:phi-simple}).  After integration over the final
position $x$ we obtain
\begin{equation}  \label{eq:rho-final}
\rho(x_0,T)  =  \frac{e^{-c_0(T)/{\sigma^2}} }{\phi(x_0,t\!=\!0)}
\int_{-\infty}^0 \ \mathrm{d}x \,
  e^{{x a_0}/{\sigma^2} } G_0^{\rm abs}(x,T|x_0)  \; .
\end{equation}

With Eq.~(\ref{eq:rho-final}) we actually obtain an {\em exact
    solution} of the first part of the program defined in
  section~\ref{sec:model}. Indeed, both $\phi(x_0,t\!=\!0)$ (cf.\
  Appendix~\ref{app:ExSol}) and the integral in the r.h.s.\ of
  Eq.~(\ref{eq:rho-final}) can be written explicitly in terms of the
  complementary error function $\operatorname{erfc}$ and elementary
  functions.  We thus have an {\em exact} and {\em explicit}
  expression for the quantity $\rho(x_0,T)$ required to address self
  consistency (cf Eq.~(\ref{eq:SCC2}) and the discussion below).
  Before we do so however, we will consider the large and small
  $\sigma$ asymptotics of this exact result; and relate them to the
  expressions obtained in section~\ref{subsec:lim}.
 
\subsection{Asymptotic regimes}

For large $\sigma$'s, and more specificaly when the condition
(\ref{cond:large-sigma}) is met, the approximation
(\ref{eq:phi-largesigma}) can be used in Eq.(\ref{eq:G-final}) and we
get
\begin{equation}
\begin{aligned}  
	G(x,T|x_0) & = e^{-\frac{1}{\sigma^2}(c(T)-c(0))}  
  e^{-\frac{1}{\sigma^2} (x_0-x) a_0} \,  G_0^{\rm abs}(x,T|x_0)  \\
&
 = e^{-\frac{1}{2\sigma^2}(c(0)-c_0(0))} G^{\rm CD}(x,T|x_0) \; .
\end{aligned}
\end{equation}	
Up to the factor $e^{-\frac{1}{2\sigma^2}(c(0)-c_0(0))} = 1 +
O(\sigma^{-2})$, one  thus recovers the 
elementary solution of the convection-diffusion
Eq.~(\ref{eq:Green-CD})  so that 
\begin{equation}
\begin{aligned} \label{eq:rho-large-sigma}
	\rho(x_0,T) & \simeq \int_{-\infty}^0 G^{\rm CD}(x,T|x_0) \
        \mathrm{d}x \\
        & = \frac{1}{2} \left[ \operatorname{erfc}\left(\frac{x_0+a_0
            T}{\sqrt{2 \sigma^2 T}}\right)-e^{-2 a_0 x_0 / \sigma^2 }
          \operatorname{erfc} \left(-\frac{x_0-a_0 T}{\sqrt{2 \sigma^2
                T}}\right) \right] \; . 
\end{aligned}
\end{equation}
As we shall see in section~\ref{sec:SCR}, this equation will be mainly
useful in the diffusion regime where, beyond the condition
(\ref{cond:large-sigma}) one may assume $x_0^2 \ll \sigma^2 T$. In
this case Eq.~(\ref{eq:rho-large-sigma}) simplifies to 
\begin{equation} \label{eq:rho-large-sigma-diff}
\rho(x_0,T) = - \frac{2 x_0}{\sqrt{2\pi \sigma^2 T}} \exp \left( -
  \frac{a_0^2 T}{2 \sigma^2} 
  \right)\; . 
\end{equation}

\begin{figure}[ht]
\centering
\includegraphics[width=10cm]{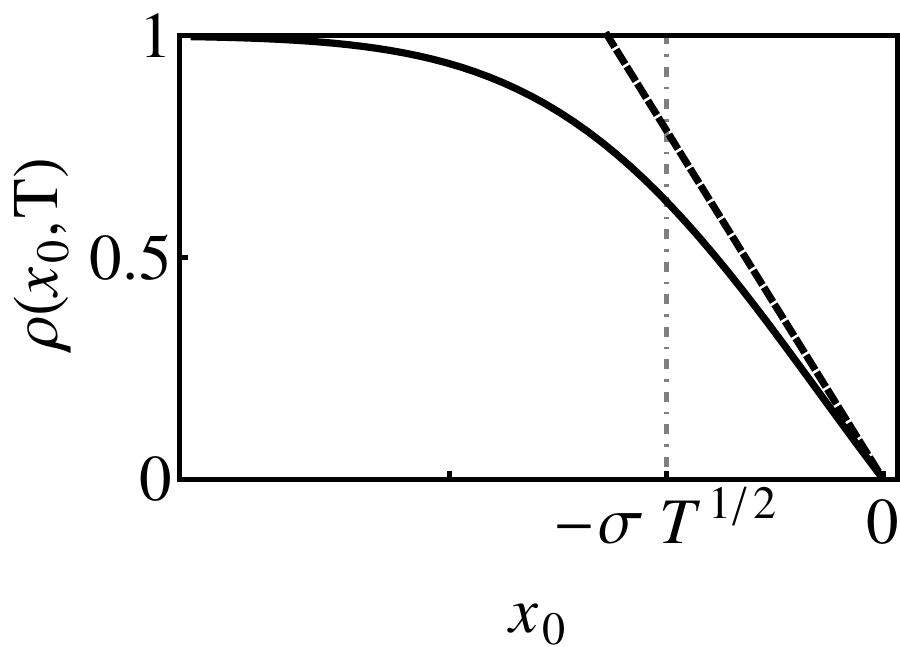}
\caption{  Solid line: asympotic form
    Eq.~(\ref{eq:rho-large-sigma}) of $\rho(x_0,T)$ valid for large
    $\sigma$'s (i.e.\ under the condition (\ref{cond:large-sigma}))
    and for $T$ such that ($a_0 T \ll \sqrt{\sigma^2 T}$); Dashed :
    linear behaviour Eq.~(\ref{eq:rho-large-sigma-diff}) corresponding
    to the diffusion regime. [$\alpha =2$, $\beta=1$, $T=1.1$ and
    $\sigma^2=49$].  }
\label{fig:rho-sketch}
\end{figure}

Another asymptotic regime is obtained when the diffusion time
$t_\sigma = x_0^2 / \sigma^2$ is much larger than the drift time
$t_{\rm d}(x_0)$ (defined as the arrival time for the participant
located at $x_0$ when $\sigma=0$; thus here $t_{\rm d} = |x_0|/a_0$ in
regions (0) $t_{\rm d} = |x_0|/a_2$ in regions (2) $t_{\rm d} = T$ in
region (1) and $t_{\rm d} = \bar t$ in region (3)).  For small
$\sigma$'s this condition applies for most of the $x_0$ axis, except a
boundary layer in the region $ x_l \leq x \leq 0$, where $x_l$ is
defined by $\sigma^2 \gg \displaystyle \frac{x_l^2}{\bar{t}}$.

\begin{figure}[ht]
\centering
\includegraphics[width=10cm]{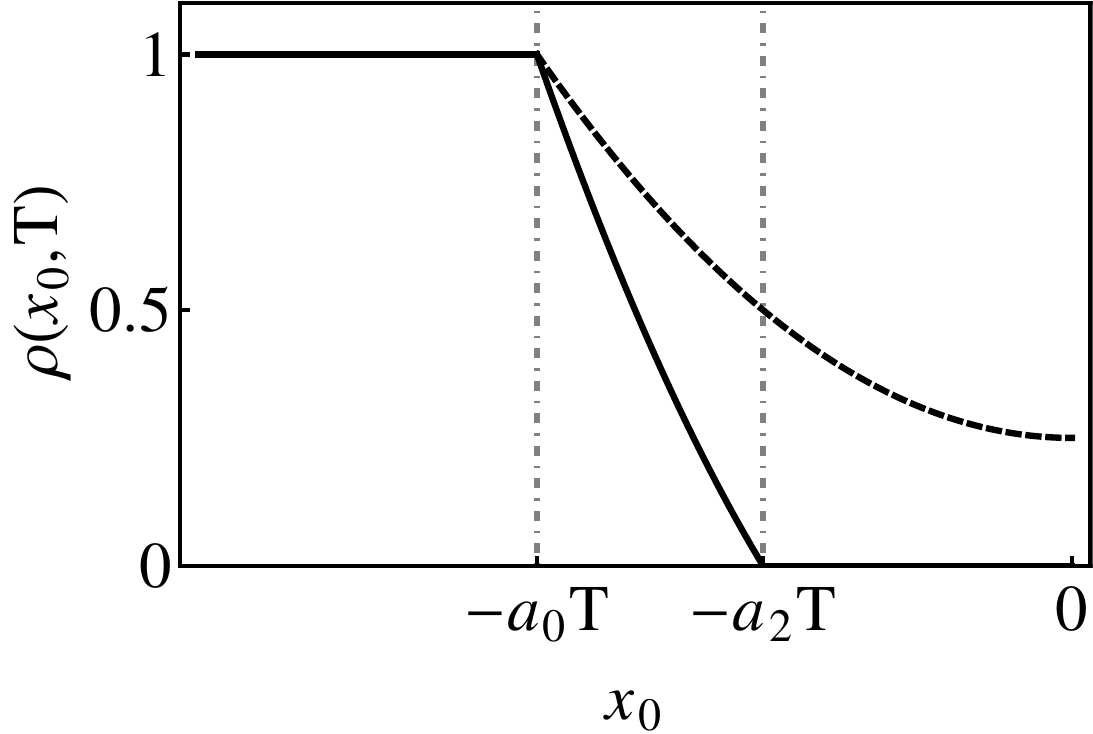}
\caption{Asympotic form of $ \rho(x_0,T)$ for $\sigma$ small.
    Full line: $T>\bar{t}$; dashed: $T=\bar{t}$.
[$\alpha =2, \
    \beta=1, \ \gamma=1, \ T=2 $].  
}
\label{fig:rho-conv}
\end{figure}

In this asymptotic regime the Laplace method (\cite{WongBook}) can be
used to evaluate the integrals occuring in Eq.~(\ref{eq:rho-final})
(we detail the calculation of $\phi(x_0,0)$, in
appendix~\ref{app:phi-smallsigma}, and the evaluation of the
 numerator is done along the same lines). We
get
\begin{equation} \label{eq:rho-small-sigma}
\rho(x_0,T)=
\left\{
\begin{array}{cl}
1, \; &\mbox{ if } x_0 \leq -T a_0\\
\displaystyle \frac{ \frac{x^2_0}{T^2} -a^2_2}{a^2_0-a^2_2}
 \; &\mbox{ if } -T a_0 \leq x_0 \leq
-T a_2 \\ 
  0, \; &\mbox{ if } x_0  \geq -T a_2
\end{array}
\right. \; .
\end{equation}
We see in this way how the singular behaviour of the strict $\sigma=0$
limit is regularized for small but non-zero sigma (cf
Eq.~(\ref{eq:m0_small_sigma})  and the discussion below).
An illustration of this function is given on
Fig.~\ref{fig:rho-conv}.

As discussed in appendix~\ref{app:phi-smallsigma},
Eq.~(\ref{eq:rho-small-sigma}) requires to be valid that $T$ is
sufficiently distant from $\bar t$ (cf
Eq.(\ref{cond:small-sigma-1c}) for the precise condition).  If for
instance $T=\bar t$ one would have instead

\begin{equation} \label{eq:rho-small-sigma-bis}
\rho(x_0,T\!=\!\bar t)=
\left\{
\begin{array}{cl}
1, \; &\mbox{ if } x_0 \leq - \bar t a_0\\
\displaystyle \frac{ \frac{x^2_0}{2 \bar t^2} -c'_3}{c'_0-c'_3}
 \; &\mbox{ if } -\bar t a_0 \leq x_0 \leq 0
\end{array}
\right. \; .
\end{equation}
with $c'_0 = (\alpha + \beta)$ and $c'_3 = - \gamma$ the slopes of
$c(\tau)$ for $\tau > T$ and for $\tau < \bar t$ respectively.  The
shape of this function is illustrated as a dashed line in
Fig.~\ref{fig:rho-conv}.  Note that
  Eq.~(\ref{eq:rho-small-sigma-bis}) is valid only for $|x_0|$ large
  enough that the motion is dominated by convection, and in particular
  does not apply at $x_0=0$, where in any case one should have
  $\rho(x\!=\!0 , t) \equiv 0$.  The condition of validity of
  Eq.~(\ref{eq:rho-small-sigma-bis}) can be shown for small $\sigma$'s
  to read $ |x_0| \gg (a_0^2 \bar t^{3/2}/\sigma) \exp(-\gamma \bar
  t/\sigma^2)$.

\section{Self-consistent condition}
\label{sec:SCR}

It is time now to answer the question: ``when does the meeting
  start ?''.  Answering this question implies taking into account the
  coupling between agents mediated by the mean-field condition, and
  means in practice solving the self-consistent
  equation~(\ref{eq:SCC2}) with the form of the function $\rho(x_0,T)$
  given by Eq.~(\ref{eq:rho-final}).

One thing worth noticing already is that having an explicit expression
for $\rho(x_0,T)$ could provide an alternative route -- to the
one given by Gu\'eant et al. \cite{GueantLasryLions2010} -- for the
proof of the existence of the solution for $T$, which is associated
with the continuity of $\rho(x_0,T)$. This route is of course
restricted to particular models such as the present one.  
In this section
however, we are not so much interested in this ``proof of existence''
than into a  qualitative description, and 
whenever possible a more
quantitative one, of the behavior of $T$ as a function of the various
parameters of the problem.

A stressed in section~\ref{sec:model}, among these parameters the
initial density of agents $m_0(x_0)$ plays a specific role.  Indeed,
the other parameters, namely $(\alpha,\beta,\gamma)$ characterizing
 the cost function $c(t)$, the official time of seminar
$\bar t$, and the intensity of the noise $\sigma$, enter through the
function $\rho(x_0,T)$, and their specific role has been discussed at
length when analyzing the property of this function.  The initial
density $m_0(x_0)$ on the other hand enters only now in the
discussion  since the behavior of the agents is coupled only through $T$. 
Furthermore, $m_0(x_0)$ being a function, it may have a
infinite variety of shape, and it is clearly not realistic to discuss
the more esoteric among them.  In the following, we shall therefore
restrict our study to initial distributions $m_0(x_0)$ that can be
characterized by their mean value $\langle x_0 \rangle$ and their
variance $\Sigma^2$, and thus implicitly assume that  $\Sigma$ sets a
scale below which the variations of $m_0(x_0)$ are small.

The mean value $\langle x_0 \rangle$ will mainly determine how much
$T$ is influenced, or not, by the official time of the seminar.
Clearly, if $\langle x_0 \rangle$ is close enough to zero,  almost
  all the mass will be close to the origin of the negative semiaxis
  and there is a point where the noise $\sigma$ will be sufficient to
fill the seminar room, and  the quorum will
be met before the official beginning time, giving $T = \bar t$. For
larger, but not too large $|\langle x_0 \rangle|$, agent have a real
possibility to arrive in the seminar room near, or even a bit before,
$\bar t$, which will influence their optimization choices, and
eventually lead to a self consistent $T$ which depend on $\bar t$,
although $T > \bar t$.  For very negative $\langle x_0 \rangle$ on the
other hand, there is very little chance for an agent to arrive before
$\bar t$.  Indeed, as we have seen in section~\ref{sec:HJB}, it is
never optimal for a agent to choose a drift velocity higher than $a_0
= \sqrt{2c'_0}$, where $c'_0$ is the slope of the cost function $c(t)$
for time $t \geq T$. As a consequence, if $|\langle x_0 \rangle| \gg
a_0 \bar t$, the agents determining the quorum condition (ie the last
ones to arrive before the quorum is met) will never consider the
possibility to arrive before $\bar t$, and thus the official starting
time of the seminar will play no role in setting $T$.

The parameter $\Sigma$ on the other hand will balance the effect of
the  Kolmogorov  diffusion term  in 
the determination of $T$.  Indeed, a set of agents starting from an
identical initial location will have spread on a distance
$\sigma\sqrt{T} $ at time $T$.  So for 
\begin{equation} \label{cond:narrow}
\Sigma \ll \sigma\sqrt{T}
\end{equation}
 the diffusion will essentially erase any of the initial features of
$m_0(x_0)$,  while  for 
\begin{equation} \label{cond:wide}
\Sigma \gg \sigma\sqrt{T}
\end{equation}
diffusion plays little role for the transport of $m(x,t)$.  Keeping in
mind this general picture, we turn now to a more detailed description
of the various limiting cases.

\subsection{Self consistent condition in the diffusion regime (large
  \texorpdfstring{$\sigma$'s}{sigma's}) }

We characterize the diffusion regime by the fact that $\sigma$ is
large enough (condition (\ref{cond:large-sigma})) and 
that, for the relevant positions $x_0$, 
the time of drift is much larger than the diffusion time, i.e. here
\begin{equation} \label{cond:diff-1}
 |x_0| a_0 \ll \sigma^2  \ ; .
\end{equation}
We consider successively  narrow initial distributions and wider ones.

\subsubsection{Narrow initial distributions}

A narrow initial distributions corresponds to a configuration where
the initial width $\Sigma$ is significantly smaller than the spreading
$\sigma \sqrt{T}$ acquired because of the noise during the transport --
 the notion of narrow initial distribution is thus $\sigma$-dependent,
and this configuration will typically be met when the noise is rather
large.  In that case the details of the initial distribution become
irrelevant and $m_0(x_0)$ can be approximated by a Dirac function
$\delta(x_0 -\langle x_0 \rangle)$).  The integral equation
(\ref{eq:SCC2}) therefore just become the simple standard equation
\begin{equation} \label{eq:SCC-ls}
\rho(\langle x_0 \rangle,T) = \bar \theta \; ,
\end{equation}
in which $m_0(x_0)$ is entirely characterized by its mean value
$\langle x_0 \rangle$.
In the  large noise regime  the conditions
(\ref{cond:large-sigma}) and (\ref{cond:diff-1}) hold.  Furthermore, we
will see that for the self consistent value of $T$ obtained at the end
of the process one gets
\begin{equation} \label{cond:large-sigma-diff}
  \langle x_0 \rangle^2 \ll \sigma^2 T  \ ; .
\end{equation}
Under these conditions, we can use 
the approximation Eq.~(\ref{eq:rho-large-sigma-diff})  for
$\rho(x_0,T)$, and  Eq.~(\ref{eq:SCC-ls}) reads 
$e^{-u} /\sqrt{\pi u} = \bar \theta \sigma^2 / (a_0 \langle x_0
\rangle)^2$, with $u \equiv a_0^2 T / 2 \sigma^2$.

If $\bar \theta \gg a_0 |\langle x_0\rangle| / \sigma^2 $, which deep
in the diffusive regime will usually holds except for very small $\bar
\theta$, we get in leading $1/\sigma$ order, $T = \sup(T^*, \bar t)$,
with
\begin{equation} \label{eq:T-LN-Ls-close}
T^* \simeq \frac{2}{\pi} \frac{\langle x_0 \rangle^2}{\sigma^2 \bar
  \theta^2} \; .
\end{equation}
(If $T = T^*$, the condition (\ref{cond:large-sigma-diff}) then just
amounts to have $\bar \theta \ll 1$, which we assume.  If $T = \bar
t$, the condition (\ref{cond:large-sigma-diff}) is even more easily
fulfilled.)  $T^*$ is proportional to $\langle x_0 \rangle^2$, and we
recover the intuitive result that if the initial distribution is
located too close from the seminar room, the noise fills this latter
before the official starting time, giving $T=\bar t$.

For very small $\bar \theta$, there is a range of $\sigma^2$ for which
even in the diffusive regime (\ref{cond:diff-1}) one has $\bar \theta
\ll a_0 |\langle x_0\rangle| / \sigma^2 $.  In that case
\begin{equation} \label{eq:T-LN-Ls-close-bis}
T \simeq \frac{2\sigma^2}{a_0^2} \log \left( 
\frac{a_0|\langle x_0 \rangle|}{\sqrt{\pi}\sigma^2 \bar
  \theta} \right) \; ,
\end{equation}
and one can check that (\ref{cond:large-sigma-diff}) holds.

\subsubsection{Wide initial distributions}

When $\Sigma \gg \sigma \sqrt{T}$ -- which since we assume here
$\sigma^2 \gg (c(0)-c_0(0))$ implies fairly large $\Sigma$'s -- the
convolution with a Gaussian of width $\sigma \sqrt{T}$ barely change
the distribution.  Everything appear then as if the Kolmogorov
approximation was dominated by convection.

Let us introduce $x_{\theta}$ such that
\begin{equation} \label{eq:xtheta}
\int_{-\infty}^{x_\theta} m_0 (x_0) dx_0 =  \bar \theta\; ,
\end{equation}
which is thus the position of the participant such that  a
  fraction  $\bar \theta$ of the agents is more distant from the
  origin.  The beginning of the seminar is entirely determined by the
time at which the agent starting from this location and evolving in a
deterministic way under the influence of the drift $a(x,t)$ (ie
ignoring the effect of the noise) will arrive.

In the large noise limit that we consider here, the drift is constant
and equal to $a_0$, and this just gives 
\[ T= - \frac{x_{\theta}}{a_0} \; . \]
(The fact that $x_\theta$ is necessarily of the order of or larger
than $\Sigma$, together with (\ref{cond:large-sigma}), implies $T >
\bar t$.)

\subsection{Self consistent condition in the convection regime}

In the convection regime, and more precisely under the
conditions~(\ref{cond:small-sigma-1a})-(\ref{cond:small-sigma-1c}),
the function $\rho(x,T)$ is well approximated by
Eq.~(\ref{eq:rho-small-sigma}).  We consider below how
Eq.~(\ref{eq:SCC2}) can be solved for this form of $\rho(x,T)$ for
different ranges of the initial distribution's width $\Sigma$.

\subsubsection{Narrow initial distributions}

For very narrow initial condition, Eq.~(\ref{eq:SCC2}) can be as before
replaced by Eq.~(\ref{eq:SCC-ls}) which, with
Eq.~(\ref{eq:rho-small-sigma}), is solved  as
\begin{equation} \label{eq:T-SS-Ss}
\begin{aligned} 
T &= \frac{|\langle x_0 \rangle|}{\bar{a}(\bar \theta)} \\
\bar{a}(\bar \theta) & \equiv {\sqrt{a_2^2 + (a_0^2 - a_2^2)\bar
    \theta}} \; . 
\end{aligned}
\end{equation}  
For small $\bar \theta$, Eq.~(\ref{eq:T-SS-Ss}) corresponds to
$\bar{a}(\bar \theta) \simeq a_2$, ie to a $\langle x_0 \rangle$ near
the lower border of region(1), for which the
condition~(\ref{cond:small-sigma-1b}) might not be fulfilled.
Re-inserting Eq.~(\ref{eq:T-SS-Ss}) into (\ref{cond:small-sigma-1b})
we indeed see that Eq.~(\ref{eq:T-SS-Ss}) applies only if
\begin{equation}
     \sigma^2 \ll a_0 |\langle x_0 \rangle | \bar \theta^2 \; .  
\end{equation}

For larger $\sigma$'s -- or smaller $\bar \theta$ -- we need to use
for $\phi(x,0)$ in Eq.~(\ref{eq:rho-final}) the uniform approximation
Eq.~(\ref{eq:phi-uniform}) valid for $x$ near $-a_2 T$.  Writing $T =
T_0 + \delta T$ with $T_0 = |\langle x_0 \rangle |/a_2$, we find then
in linear order
\begin{equation} \label{eq:deltaT-uniform}
\frac{\delta T}{T_0} = \frac{\sqrt{\pi}}{2} \frac{a_0^2 - a_2^2}{a_2^2} (\bar
\theta_0 - \bar \theta) \; ,
\end{equation}
with $\bar \theta_0 \equiv \sqrt{8 \sigma^2 / \pi  |\langle x_0
  \rangle | a_2} (a_2^2/(a_0^2 - a_2^2))$.

The expressions (\ref{eq:T-SS-Ss})-(\ref{eq:deltaT-uniform}) are
clearly independent of the official beginning time of the seminar
$\bar t $. The condition (\ref{cond:small-sigma-1c}), which is
actually required for Eq.~(\ref{eq:rho-small-sigma}) to apply, indeed
implies that $T$ is sufficiently above $\bar t$ to become independent
of this latter.

Once $|\langle x_0 \rangle|$ diminishes, and more specifically  when 
it reaches a value close to $a_2 \bar t$ or smaller, $T$ will on the
other hand approach $\bar t$.  It may be interesting then to determine
under which condition one has exactly $T= \bar t$, i.e.\ when the
quorum is met {\em before} the official beginning time $\bar t$.

In the convection regime,   $\rho(x,T\!=\! \bar t)$ is
described by the expression Eq.~(\ref{eq:rho-small-sigma-bis}), and
the self consistent condition to obtain $T = \bar t$ is that
  \begin{equation} \label{eq:TisTbar}
 \rho(\langle x_0 \rangle,T\!=\! \bar t) < \bar \theta \; .
\end{equation}
The approximation Eq.~(\ref{eq:rho-small-sigma-bis}) is however
bounded from below by its value at zero, $\gamma /
(\alpha+\beta+\gamma)$, which is typically of order one.  If, as we assume,
$\bar \theta$ is small, Eq.~(\ref{eq:TisTbar}) will thus not have any
solution for $\langle x_0 \rangle$ is the convection regime.  As long
as the motion of the agents is convective, they will manage to fill
the seminar room  after, though possibly  barely, the official start of the
seminar. 

If $| \langle x_0 \rangle|$ becomes so small that the motion at such
distance is dominated by diffusion, then again one can eventually
reach a point where the quorum is met before $\bar t$, implying
$T=\bar t$.  For small $\sigma$'s it can be shown that this happens
when $| \langle x_0 \rangle| \simeq \sqrt{\pi/2} (a_0^2\bar
t^{3/2}/\sigma) (\bar \theta /(1 -\bar \theta))  \exp(-\gamma \bar
t/\sigma^2)$.

\subsubsection{Wider initial distributions}

If the width of the initial distribution is non negligible, we need to
distinguish two cases.  For intermediate values of $\Sigma$, namely
for $\Sigma$'s such that once self-consistence is obtained most of the
initial distribution is in the range $]-a_0 T , -a_2 T[$, we can use
that in this range the function $\rho(x,T) \simeq (x^2/T^2 -
a_2^2)(a_0^2 - a_2^2)$ is a simple polynomial.  The convolution with
$m_0(x)$ thus simply leads to
\[ 
\int_{-\infty}^0dx_0 \, \rho(x,T) m_0(x) =  \frac{(\langle
  x_0\rangle^2 + \Sigma^2)/T^2 - a_2^2}{a_0^2 - a_2^2} \; ,
\]
and  Eq.~(\ref{eq:T-SS-Ss}) has just to be replaced by 
\begin{equation} \label{eq:T-SS-Ss2}
T =\sqrt{ \frac{\langle x_0 \rangle^2 +\Sigma^2 }{a_2^2 + (a_0^2 - a_2^2)\bar
    \theta}} \; . 
\end{equation}
The constraint that the initial distribution fits within $]-a_0 T ,
-a_2 T[$ implies that Eq.(\ref{eq:T-SS-Ss2}) applies only if $\Sigma
\ll \bar \theta |\langle x_0 \rangle|$, i.e.\ for not too small $\bar
\theta$.  For smaller $\bar \theta$, explicit (but less transparent)
expressions can be written down under the less restrictive condition
$\Sigma < (a_0 - a_2) T \simeq ((a_0 - a_2)/a_2) |\langle x_0 \rangle
|$ for specific forms of the initial distribution (eg Gaussian).

If now $\Sigma$ is large not only on the scale $\sigma \sqrt{T}$ but
also on the scale $(a_0 - a_2) T$, another approach can be used.
Subtracting Eq.~(\ref{eq:xtheta}) to Eq.~(\ref{eq:SCC2}) and
neglecting the variation of $m_0(x_0)$ near $x_\theta$ in the whole
region (1), we can write that $\int_{-a_0 T}^{-a_2 T}
\frac{x_0^2/T^2 - a_2^2}{a_0^2 - a_2^2} dx_0 = \int_{-a_0
  T}^{x_\theta} d x_0 $, which implies
\[ 
T = - \frac{3}{2} |x_\theta| \left( \frac{a_0^2 - a_2^2}{a_0^3-a_2^3}
\right) \; .
\]
As before, this results apply only if $T - \bar t$ is large enough for
Eq.~(\ref{cond:small-sigma-1c}) to be fulfilled.

\subsection{``Phase diagram'' of the seminar problem}
\label{section:phase_diag}

\begin{figure}[htb]
\includegraphics[width=10cm]{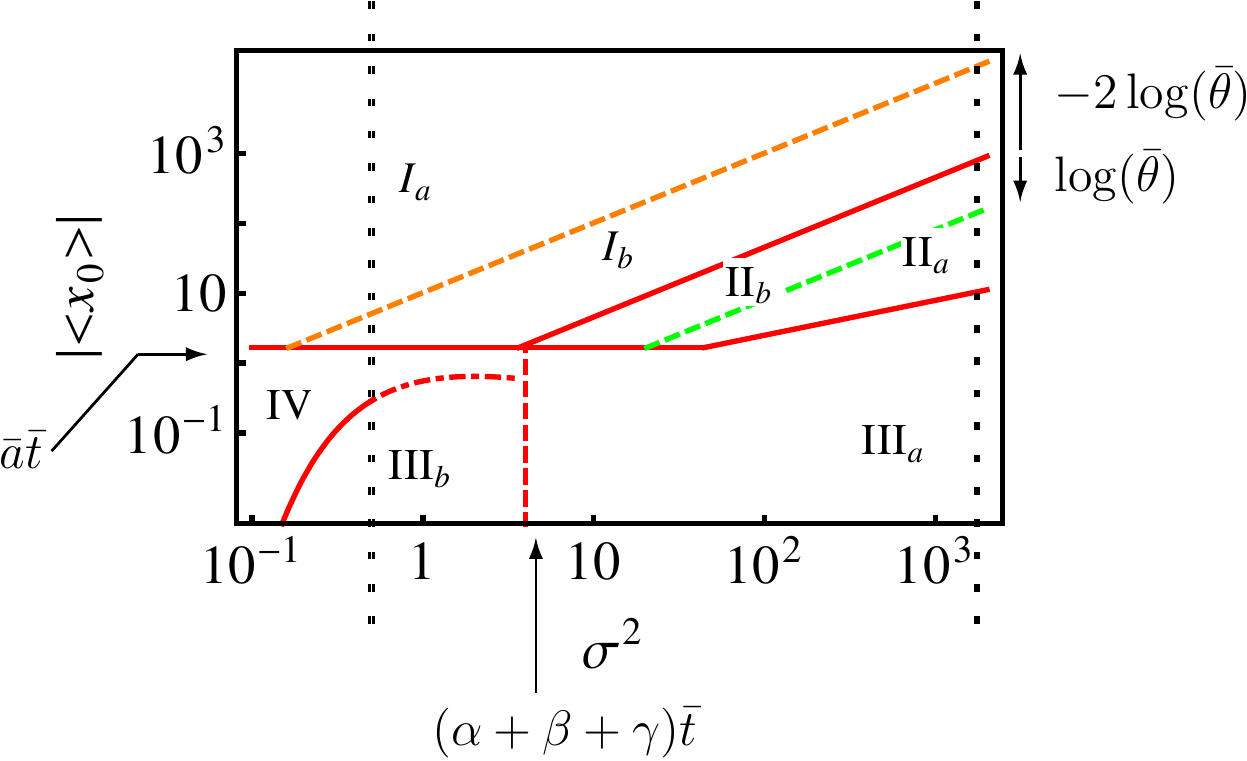}
\caption{Phase diagram of the seminar problem, in the $(\sigma^2,
  |\langle x_0 \rangle|)$ plane, for narrow initial distributions (see text for the
  detailed description of the various regimes).  The  vertical
  dashed-dotted line correspond to the vertical cuts used in
  Figs.~\ref{fig:cut_s} and \ref{fig:cut_S}.  For this illustration,
  the parameters of the problems have been  taken as $\left[ \alpha =2,
    \beta=1, \gamma=1, \bar{t}=1, \bar \theta=0.2 \right] $ }
\label{fig:Phase_Diagram}
\end{figure}

The results of the previous subsections can be summarized into
  ``phase diagrams'' such as the one shown on
  Fig.~\ref{fig:Phase_Diagram} for narrow initial distributions
  (equivalent phase diagrams can be constructed in the same way for
  wider initial distributions).  Keeping in mind that, except for the
  transition between $T = \bar t$ and $T \neq \bar t$, there is of
  course no true phase transition here, and that the lines representing the
  limits between various regimes should be though as crossover regions
  (thus with a finite extension), we can distinguish the following
  ``phases'':
\begin{itemize}
\item [\bf Region $\bf I$] corresponds to a motion dominated by convection, and
  such that the initial distribution is far enough from the seminar
  room that the initial time of the seminar becomes irrelevant.  This
  region is split into two subregions.  In the first one, $\bf I_a$,
   $T = |\langle x_0 \rangle|/\bar{a}(\bar  \theta)$ with
   $\bar{a}(\bar \theta)$ defined by Eq.~(\ref{eq:T-SS-Ss}). 
 In the second  one, $\bf I_b$, the fact that $\bar \theta \ll 1$ and
 thus that $\bar{a}(\bar  \theta) \simeq a_2$ makes it necessary to
 use the  uniform approximation Eq.~(\ref{eq:phi-uniform})
   for $\phi(\langle x_0 \rangle,T)$.  In that case
  $T={|\langle x_0 \rangle|}/{a_2}+ 
  \delta T$ where $\delta T$ is given by Eq.~(\ref{eq:deltaT-uniform}).
\item[\bf Region $\bf II$] corresponds to a motion dominated by
  diffusion, and such again that the initial distribution is far
  enough from the seminar room that the initial time of the seminar is
  irrelevant.  Region $\bf II$, too, has to be divided in two
  subregions.  In $\bf II_a$, $T = T^*$ with $T^*$ given by
  Eq.~(\ref{eq:T-LN-Ls-close}). In $\bf II_b$ the smallness of $\bar
  \theta$ should be taken into account, leading to
  Eq.~(\ref{eq:T-LN-Ls-close-bis}).
\item [\bf Region $\bf III$] corresponds to a motion which can be
  dominated either by diffusion (region $\bf III_a$) or by convection
  (region $\bf III_b$), but such that in any case the quorum is met
  before the official beginning time of the seminar.  This region thus
  correspond to the phase   $T=\bar{t}$.
\item [\bf Region $\bf III$] corresponds finally to a configuration
  such that the quorum is met slightly after the official beginning
  time of the seminar, so that $T$ is different from, but close to, $\bar{t}$.
\end{itemize}

\begin{figure}[htb]
\includegraphics[width=10cm]{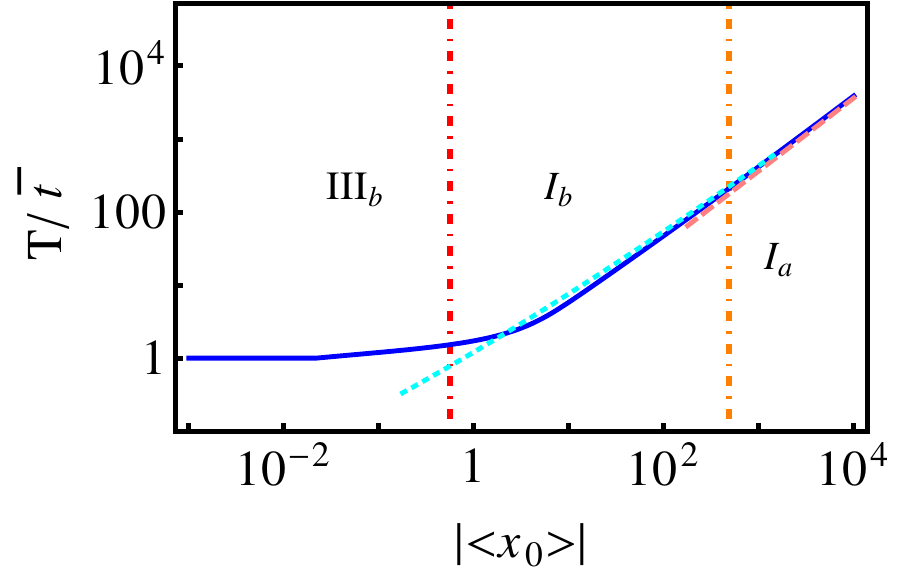}
\caption{$|\langle x_0 \rangle|$ dependence of the time $T$ for narrow initial
  distributions and a small value of the noise parameter ($\sigma =
  0.6$); which corresponds to the left vertical dashed-dotted line in
  the phase diagram Fig.~\ref{fig:Phase_Diagram}. The numerical labels
  correspond to those of the different regimes in
  Fig.~\ref{fig:Phase_Diagram}.  Full line~: numerical value obtained
  from the exact expression Eq.~(\ref{eq:rho-final}). Dashed :
  asymptotic expressions in the corresponding regime (see text).  The
  parameters of the model are the same as in
  Fig.~\ref{fig:Phase_Diagram}, except for $\bar{\theta} = 0.1$ which,
  to enhance readability, has been slightly decreased.}
\label{fig:cut_s}
\end{figure}

\begin{figure}[htb]
\includegraphics[width=10cm]{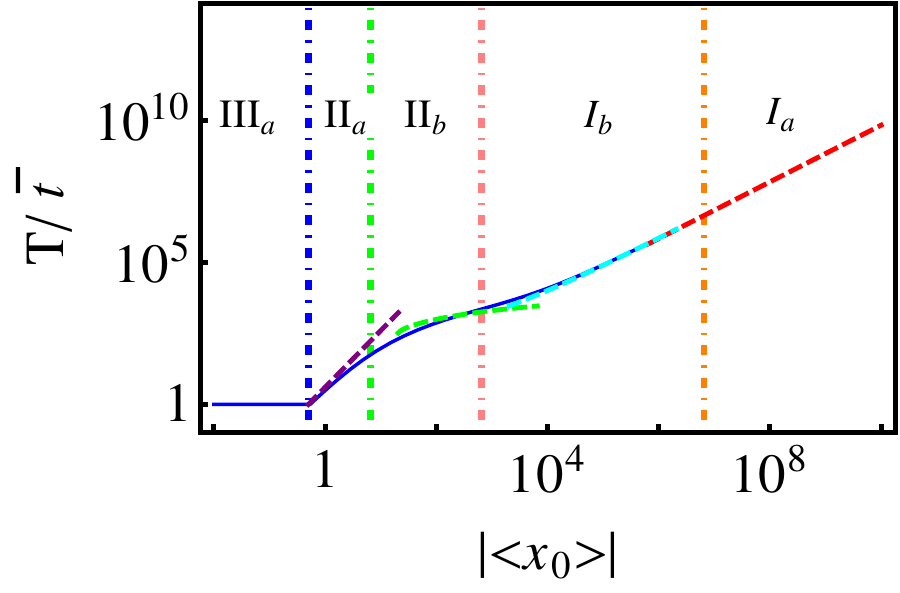}
\caption{$|\langle x_0 \rangle|$ dependence of the time $T$ for narrow initial
  distributions and a large value of the noise parameter ($\sigma =
  40$); which corresponds to the right vertical dashed-dotted line in
  the phase diagram Fig.~\ref{fig:Phase_Diagram}. The numerical labels
  correspond to those of the different regimes in
  Fig.~\ref{fig:Phase_Diagram}.  Full line~: numerical value obtained
  from the exact expression Eq.~(\ref{eq:rho-final}). Dashed :
  asymptotic expressions in the corresponding regime (see text). The parameters
  of the model are the same as in Fig.~\ref{fig:Phase_Diagram},
  except for $\bar{\theta} = 0. 01$.}
\label{fig:cut_S}
\end{figure}

As an illustration, we show in Fig.~\ref{fig:cut_s} and
\ref{fig:cut_S} two vertical cuts in this phase diagram, in which are
displayed the variations of the self-consistent time $T$ as a function
of $\langle x_0 \rangle $ for two (fixed) values of the noise
$\sigma$, one ``small'' and one ``large''. For the small diffusion
coefficient case Fig.~\ref{fig:cut_s} we observe, as expected from the
phase diagram, a transition between a domain where $T=\bar{t}$ and a
domain where $T={|\langle x_0 \rangle|}/{\bar{a}(\bar \theta)}$. For
the large $\sigma$ case Fig. \ref{fig:cut_S}, we observe, again as
predicted from the phase diagram, a richer behavior, with the same
limiting behaviors for very large and very small $|\langle x_0
\rangle|$, but a larger number of intermediate regimes.

\section{Conclusion}
\label{sec:conclusion}

In this article, we have considered a simple toy-model, sharing many of
the characteristic features of generic mean field games, but with the
essential simplification that the ``mean field'' actually reduces to a
simple number (the actual starting time of the seminar $T$).  The
study of this problem can then be divided in two essentially
independent parts~: on one hand the resolution, for arbitrary $T$, of
the system of partial differential equations
Eqs.~(\ref{eq:HJB})-(\ref{eq:KMG}) describing the coupling between the
agents optimization decisions and their motion; and on the other hand
the self-consistent condition Eq.~(\ref{eq:SCC2}) determining the
value of $T$.

The first part of this program can be performed essentially
completely.  Indeed an explicit expression Eq.~(\ref{eq:rho-final})
can be obtained on a very general basis for the function $\rho(x_0,T)$
required to discuss self-consistency.  From this general result
transparent asymptotic expressions are derived in the relevant
limiting regimes (cf e.g.\ Eqs.~(\ref{eq:rho-large-sigma}),
(\ref{eq:rho-small-sigma}) and (\ref{eq:rho-small-sigma-bis})). From
these and the self-consistent condition Eq.~(\ref{eq:SCC2}) the
qualitative behavior of $T$, and  explicit expressions in many
limiting regimes of interest, can be obtained.

We will not re-list here the various results derived for $T$ in
section~\ref{sec:SCR}, but the main features may be summarized as
follow.  An important point is that the the slopes $c'_0 = (\alpha +
\beta)$ and $c'_2 = (\alpha - \gamma)$ of the cost function $c(t)$ (cf
Eq.~(\ref{eq:c(t)})) can be associated with drift velocities $a_0 =
\sqrt{2 c'_0}$ and $a_2 = \sqrt{2 c'_2}$ which fixes the scale of the
drift velocities of the problem.  Together with the characteristic
length $l_0$ associated with the initial distribution of agent
$m_0(x_0)$ (namely the center of mass, i.e. $l_0 = |\langle x_0
\rangle|$ for narrow distributions, or $l_0 = |x_\theta|$ (cf
Eq.~(\ref{eq:xtheta})) for wide distributions) and the value of the
noise parameter $\sigma$, they organize the system ``phase diagram''
of the problem, and define the relevant limiting regime.  For instance
if $ |l_0| \gg a_0 \bar t $ and $|l_0| \gg \sigma \sqrt{\bar t}$
the system is completely dominated by convection, and $T \simeq l_0/a$,
with $a$ equal to  $a_0$ for large noise and to some weighted
average between $a_0$ or $a_2$ for smaller noise.  Or if $l_0 \ll a_2
\bar t$ one is essentially guaranteed that the quorum will be met
before the official starting time $\bar t$ of the seminar, and $T =
\bar t$, etc..

One point worth being stressed however is that in most circumstances,
either the agents start from a location close from the seminar
room ($l_0 \ll a_{0,1} \bar t$) and $T = \bar t$, or they are
initially far from the seminar room ($l_0 \gg a_{0,1} \bar t$) and $T$
becomes relatively quickly independent of $\bar t$.  The transition region
for which $T > \bar t$ but keep some $\bar t$ dependence is actually
rather restricted.  This notion of a effective starting time of the
seminar which is independent of its official starting time is clearly
a bit disturbing, especially from the viewpoint of the seminar
organizer. 

This feature can be tracked back to the fact that there exist an
initial time $t=0$ at which the all agents start their optimization
process and motion, and what we see is that this initial time plays a
role which is at least as important as $\bar t$ in the determination
of $T$.  One can of course imagine that this initial time has some
physical signification (e.g.\ time at which the organizers ring a
bell, etc..).  One could also modify slightly the model to remove the
reference to a uniform initial time and have the time $\tau_0$ at
which a given agent leave her office taken as a parameter entering in
the optimization decision.  Assuming the marginal cost of not being
in ones office is described by some parameter $\eta$, this would amount
to replacing the cost function Eq.~(\ref{eq:total_cost}) by 
\begin{equation*} 
  J_T[a]=\mathbb{E} \left [ c(\tilde{\tau}) - \eta \tau_0 + \frac{1}{2}
    \int_{\tau_0}^{\tilde{\tau}} a^2_i(\tau) \ \mathrm{d}\tau \right]
  \; .
\end{equation*}
The problem could be analyzed along the line of what we have done in
this paper, and would lead to a stronger dependence of $T$ in $\bar t$.

More generally, various variations of the problem can be easily
studied with the approach followed in this paper.  In particular,
Eqs.~(\ref{eq:G-final})-Eqs.~(\ref{eq:phi-general}) are valid for
essentially arbitrary cost functions $c(t)$, and could be studied for
instance in circumstances for which the self consistent condition
Eq.~(\ref{eq:SCC2}) has more than one solution.  The seminar problem
is therefore a very versatile model, and the very thorough
understanding of its behavior obtained in this work should help
develop the intuition on the properties of more generic mean field
game models.

\begin{acknowledgments}
We thank Jean-Michel Courtault for introducing us to the
  problematic of mean field game.  Igor Swiecicki acknowledge support
  from  the Labex MME-DII.
\end{acknowledgments}

\appendix

\section{Exact Solution of HJB Equation}
\label{app:ExSol}

In this appendix we derive exact expressions 
 for $\phi (x,t)$ (see Eq.~(\ref{eq:phi-general})).

\begin{equation}
  \phi (x,t) = \frac{|x|}{\sqrt{2\pi\sigma^2}} \int_0^\infty
  \frac{d\tau}{\tau^{3/2}} 
  e^{\textstyle{-\frac{1}{\sigma^2} \bigl( c(t+\tau) +\frac{x^2}{2
        \tau}\bigr)}} 
\end{equation}
where $c(t)$ is piece-wise linear. 
\begin{equation}
c(t)= \alpha [t-\bar t]_+ +\beta [t-T]_+ +\gamma [T-t]_+
\end{equation}

Using the explicit expression for $c(t)$ and the fact that $\bar t \le
T$, one may write 
\begin{eqnarray}
\sqrt{2\pi\sigma^2} \phi (x,t) &=& |x| 
e^{\textstyle{-\frac{\gamma (T - t)}{\sigma^2} }}
\int_0^{[\bar t -t]^+} \frac{d\tau}{\tau^{3/2}}
e^{\textstyle{-\frac{1}{\sigma^2} \bigl(-\gamma \tau +\frac{x^2}{2
      \tau}\bigr)}}\nonumber\\ 
 &+& |x|  
e^{\textstyle{-\frac{\alpha (t -\bar t)+ \gamma (T - t)}{\sigma^2}}}
\int_{[\bar t -t]^+}^{[T -t]^+} \frac{d\tau}{\tau^{3/2}}
e^{\textstyle{-\frac{1}{\sigma^2} \bigl((\alpha -\gamma )\tau
    +\frac{x^2}{2 \tau}\bigr)}}\nonumber\\ 
 &+& |x| 
e^{\textstyle{-\frac{\alpha (t -\bar t)+ \beta (t - T)}{\sigma^2}}}
\int_{[T -t]^+}^\infty \frac{d\tau}{\tau^{3/2}}
e^{\textstyle{-\frac{1}{\sigma^2} \bigl((\alpha + \beta )\tau
    +\frac{x^2}{2 \tau}\bigr)}}\nonumber 
\end{eqnarray}

We define the following integral \cite{Abramowitz&Stegun}  
\begin{eqnarray}
\mathcal I (a,b,t) &=& 
2\sqrt{\frac{b}{\pi}}
\int_0^t \frac{d\tau}{\tau^{3/2}} e^{\textstyle{-a \tau -\frac{b}{\tau}}}\nonumber\\
&=&
e^{\textstyle{- 2\sqrt{a b}}}\operatorname{erfc}(\sqrt{\frac{b}{t}}-\sqrt{a t} ) 
+ e^{\textstyle{2\sqrt{a b}}}\operatorname{erfc}(\sqrt{\frac{b}{t}}+\sqrt{a t} )
\end{eqnarray}
where the function $\operatorname{erfc}(z)$ is defined on the complex 
plane, $\lim_{x\rightarrow +\infty} \operatorname{erfc}(x) =0$,
$\lim_{x\rightarrow -\infty} \operatorname{erfc}(x) = 2$. Accordingly,
$\lim_{t\rightarrow 0} \mathcal I (a,b,t)  =0$ 
and $\lim_{t\rightarrow +\infty} \mathcal I (a,b,t)  =  2
e^{\textstyle{- 2\sqrt{a b}}}$ for $a>0$ and $b>0$. 

Thus the function $\phi(x,t)$ reads
\begin{equation} \label{eq:phi-full}
\begin{aligned}
\phi (x,t) &= \frac{1}{2}
e^{\textstyle{-\frac{\gamma (T - t)}{\sigma^2} }}
\mathcal I (-\frac{\gamma}{\sigma^2},\frac{x^2}{2 \sigma^2},[\bar t -t]^+)
\\
&+ \frac{1}{2}
e^{\textstyle{-\frac{\alpha (t -\bar t)+ \gamma (T - t)}{\sigma^2}}}
\bigl(\mathcal I (\frac{a_2^2}{2\sigma^2},\frac{x^2}{2
  \sigma^2},[T -t]^+) 
- \mathcal I (\frac{a_2^2}{2\sigma^2},\frac{x^2}{2
  \sigma^2},[\bar t -t]^+) 
\bigr) 
\\
&+ \frac{1}{2} 
e^{\textstyle{-\frac{\alpha (t -\bar t)+ \beta (t - T)}{\sigma^2}}}
\bigl(
2
e^{\textstyle{-\frac{a_0 |x|}{\sigma^2}}} -
\mathcal I (\frac{a_0^2}{2\sigma^2},\frac{x^2}{2 \sigma^2},[T -t]^+)
\bigr) \; ,
\end{aligned}
\end{equation}
(where $a_0 = \sqrt{2 (\alpha +\beta)}$, $a_2 = \sqrt{2 (\alpha
    - \gamma)}$).

Eq.~(\ref{eq:phi-full}) applies for arbitrary values of time and position.
It takes however a simpler form in some time intervals:
\begin{itemize}
\item[-]
For $t \geq T$ the expression reduces to
\begin{equation}\label{eq:phi-simple}
\begin{aligned}
\phi (x,t) & = e^{\textstyle{-\frac{c(t)}{\sigma^2}}}
\exp \left(-\frac{1}{\sigma^2} {\sqrt{2 (\alpha +\beta)} |x|} \right) \\
& = 
e^{\textstyle{-\frac{\alpha (T\! -\! \bar t)}{\sigma^2}}}
\exp \left(\frac{ a_0^2 (T\!-\!t) -  a_0 |x|}{2\sigma^2} \right) 
\end{aligned}
\end{equation}
\item[-]
For $\bar t \leq t \leq T$ Eq.~(\ref{eq:phi-full}) can be   written as 
\begin{equation} \label{eq:phi-full-c}
\begin{aligned}
\phi(x,t)  & = \frac{e^{-\alpha (T-\bar t)/\sigma^2}}{2} \times
\\
& \left[ \exp \left( \frac{a_2^2(T \! - \! t) + 2a_2x}{2 \sigma^2} \right)
       \operatorname{erfc} \left(\frac{-x - a_2 (T \! - \! t)}{\sqrt{2 \sigma^2
         (T \! - \! t)}} \right) \right.  
\\
& + \exp \left(\frac{a_2^2(T \! - \! t) - 2a_2x}{2 \sigma^2} \right)
       \operatorname{erfc} \left(\frac{-x + a_2 (T \! - \! t)}{\sqrt{2 \sigma^2
         (T \! - \! t)}} \right) 
\\
& +  \exp \left(\frac{a_0^2(T \! - \! t) + 2a_0x}{2 \sigma^2} \right)
       \operatorname{erfc} \left( \frac{x + a_0 (T \! - \! t)}{\sqrt{2
         \sigma^2 (T \! - \! t)}} \right)
\\
& -  \left. \exp \left(\frac{a_0^2(T \! - \! t) - 2a_0x}{2 \sigma^2} \right)
       \operatorname{erfc} \left(\frac{- x + a_0 (T \! - \! t)}{\sqrt{2
         \sigma^2 (T \! - \! t)}} \right) 
     \right] \; .
\end{aligned}
\end{equation}

\item[-] At $t=0$ Eq.~(\ref{eq:phi-full})  reads
\begin{align}
\phi (x,0) &= \frac{1}{2} 
e^{\textstyle{-\frac{\gamma T}{\sigma^2} }}
\mathcal I (-\frac{\gamma}{\sigma^2},\frac{x^2}{2 \sigma^2},\bar t)
\nonumber\\
&+ \frac{1}{2}  
e^{\textstyle{\frac{\alpha \bar t - \gamma T}{\sigma^2}}}
\bigl(\mathcal I (\frac{\alpha -\gamma}{\sigma^2},\frac{x^2}{2 \sigma^2},T)
- \mathcal I (\frac{\alpha -\gamma}{\sigma^2},\frac{x^2}{2 \sigma^2},\bar t)
\bigr)
\label{eq:phi-t0}\\
&+ \frac{1}{2}  
e^{\textstyle{\frac{\alpha \bar t+ \beta  T}{\sigma^2}}}
\bigl(
2
e^{\textstyle{-\frac{\sqrt{2 (\alpha +\beta)} |x|}{\sigma^2}}} -
\mathcal I (\frac{\alpha + \beta}{\sigma^2},\frac{x^2}{2 \sigma^2},T)
\bigr) \; .
\nonumber
\end{align}
\end{itemize}

Finally, we stress that for large $\sigma$'s, and more precisely under
the condition $(\alpha,\beta,\gamma) \bar t \ll \sigma^2$ (but
irrespective of the value of $T$ and $x$) {\em
  Eq.(\ref{eq:phi-full-c}) provides a good approximation of the exact
  $\phi (x,0)$   for $(t \leq \bar t)$} ($t$ and $\bar  t$ can then be
set to zero in this equation).

\section{Evaluation of \texorpdfstring{$\phi(x,t)$}{phi} for large 
  \texorpdfstring{$\sigma$}{sigma}'s (diffusion regime)}
\label{app:phi-largesigma}

In this appendix, we evaluate the large $\sigma$ asymptotic of the
function $\phi(x,t)$ defined by Eq.~(\ref{eq:phi-general}).  For this
purpose, let us introduce $c_0(t) = \alpha (t - \bar t) +\beta (t -
T)$, the linear function such that $c(t) = c_0(t)$ for $t \geq T$, and
\begin{align}  
\phi_0(x,t) \equiv &   -x \int_0^{\infty}
 \frac{\mathrm{d}\tau}{\tau}
 \exp\left({-\frac{c_0(t+\tau)}{\sigma^2}}\right) G_0(x,\tau) \,
  \label{eq:phi0} \\
= & 
\exp \left( -\frac{1}{\sigma^2} [ c_0(t) + a_0|x| ] \right) 
\end{align}
(this last expression exactly corresponds to Eq.~(\ref{eq:phi-simple}),
and is obtained in the same way).

The difference between  $\phi(x,t)$ and its approximation
$\phi_0(x,t)$ can be expressed as 
\begin{equation}
|\phi(x,t) - \phi_0(x,t)| = -x \int_0^{\infty}
 \frac{\mathrm{d}\tau}{\tau}
 \exp\left({-\frac{c_0(t+\tau)}{\sigma^2}}\right) G_0(x,\tau)
 K(t+\tau)
\end{equation}
where
\[
K(\tau) \equiv  1 - e^{\displaystyle{-\frac{1}{\sigma^2} (c(\tau) - c_0(\tau))}}
\]
is a positive (since $c(\tau) \geq c_0(\tau)$) decreasing continuous
function which is uniformly zero for $\tau$ larger than $T$.  We thus
have $|\phi(x,t) - \phi_0(x,t)| \leq \phi_0(x,t) K(t)$.

As soon as 
\begin{equation}
\sigma^2 \gg (c(t) - c_0(t)) \; , 
\label{cond:large-sigma}
\end{equation}
$K(t)$ is $O(\sigma^{-2})$, and therefore 
\begin{equation} 
\phi(x,t) = \phi_0(x,t) (1+O(\sigma^{-2}) ) \; .
 \label{eq:phi-largesigma-app} 
\end{equation}

Eq.~(\ref{eq:phi-largesigma-app}) applies whenever the condition
(\ref{cond:large-sigma}) is fulfilled.   In practice however,  it is
mainly useful if the diffusive regime, i.e. when
\begin{equation} \label{cond:diff_regime}
  a_0 |x| \ll \sigma^2 
\end{equation}
($a_0 = \sqrt{2 (\alpha +\beta)}$).  In that case, since
Eq.~(\ref{cond:large-sigma}) morally implies $\sigma^2 \gg
|c_0(\tau)|$ (this is clear as soon as $t \leq (\alpha \bar t + \beta
T)/(\alpha + \beta)$ since then $c_0(t) < 0$, but remains generally
true unless $\tau \simeq T$), one has $\phi_0(x,t) =
 (1+O(\sigma^{-2}))$, and
Eq.~(\ref{eq:phi-largesigma-app}) provides little information on the
variations of $\phi(x,t)$.

It may be therefore interesting in this case to compute the $O(\sigma^{-2})$
corrections.   Noting that $K(\tau) = 0$ for $\tau > T$, we have
\begin{equation}
\phi(x,t) - \phi_0(x,t) = -x \int_0^{T-t}
 \frac{\mathrm{d}\tau}{\tau}
 \exp\left({-\frac{c_0(t+\tau)}{\sigma^2}}\right) G_0(x,\tau)
 (K(t) + \delta K(\tau)) \, , \label{eq:-largesigma-bis}
\end{equation}
where $\delta K(\tau) \equiv (K(t + \tau)- K(t))$.  The term involving
$\delta K$, which is linear in $\tau$ near $0$ (and thus do not
benefit from the $\tau^{-1/2}$ divergence)  can be shown to be
$O(\sigma^{-3})$ relative to $\phi_0(x,t)$, and we get
\begin{align}
\phi(x,t) &= (\phi_0(x,t) + K(t) )(1 +
O(\sigma^{-3})) \\
& = 
\exp \left( -\frac{1}{\sigma^2} [ c(t) + a_0 |x| ] \right) (1 +
O(\sigma^{-3}))  \; ,
\end{align}
valid therefore when both conditions (\ref{cond:large-sigma}) and
(\ref{cond:diff_regime}) apply.

\section{Evaluation of \texorpdfstring{$\phi(x,t)$}{phi} for small
  \texorpdfstring{$\sigma$}{sigma}'s (convection regime)} 
\label{app:phi-smallsigma}

In this appendix we evaluate the small $\sigma$
asymptotics of $\phi(x,t)$  using the
saddle point approximation (in regions (0) and (2)),  and more
generally the Laplace method.

Introducing 
\[
\Phi(\tau) = c(t+\tau) + \frac{x^2}{2\tau} \; ,
\]
the integral we want to compute is of the form
\[
\phi(x,t) = \frac{1}{\sqrt{2\pi \sigma^2 }} \int_0^\infty d \tau  f(\tau)
\exp\left(-\frac{\Phi(\tau)}{\sigma^2} \right) \; ,
\]
with $f(\tau) = -x / \tau^{3/2}$, and will therefore be dominated for
small $\sigma$'s by the minima of $\Phi(\tau)$.

The condition $\Phi'(\tau^*) = 0$ leads to the equation 
\begin{equation} \label{eq:saddle}
\tau^* = -x/\sqrt{2c'(t+\tau^*)}
\end{equation}
 which admits a solution in regions (0) and (2)
but not in regions (1) and (3).  One therefore has to use the saddle point
approximation in regions (0) and (2), and boundary contributions is
region (1) and (3).

\subsection{Regions (0) and (2)}

In regions (0) and (2), the stationary point is 
\begin{align*}
\tau^* &= -x/a_0  \qquad \mbox{[in region (0)]} \\
\tau^* &= -x/a_2  \qquad \mbox{[in region (2)]} \; ,
\end{align*}
which, with $\Phi''(\tau) = {x^2}/{\tau^3}$, gives within the saddle
point approximation
\begin{equation} \label{eq:phi-col}\
\begin{aligned}
\phi(x,t) & \simeq 
\exp\left(-\frac{\Phi(\tau^*)}{\sigma^2} \right) \\
& = \exp \left(-\frac{\alpha (T-\bar t)}{\sigma^2} \right)
  \exp \left( \frac{a_{0,2}^2(T \! - \! t) + 2a_{0,2}x}{2 \sigma^2} \right)\; .
\end{aligned}
\end{equation}
This approximation is valid as long as $\sigma^2 \ll |x|a_{0,2}$, or
in other words as long as the ratio between the drift time $t_d \equiv
|x|/a_{0,2}$ and the diffusion time $t_\sigma \equiv x^2 / \sigma^2$
is small.

\subsection{Regions (1)}

In region (1), there are no solution to Eq.~(\ref{eq:saddle}) as the
minima of $\Phi(\tau)$ correspond to a discontinuity of the cost
functions $c(t+\tau)$ (at $t+\tau = T$). Linearizing $\Phi(\tau)$ on
both side of this discontinuity and neglecting the variation of
$f(\tau)$ we get
\begin{equation} \label{eq:phi-small-sigma}
\begin{aligned}
\phi(x,t) \simeq  -x \sigma \sqrt{\frac{2(T\!-\!t)}{\pi}} &
\exp\left( - \frac{1}{\sigma^2} \left[ c(T) + \frac{x^2}{2(T-t)} 
  \right] \right) 
 \\
\times &
\left[ \frac{1}{a_0^2 (T\!-\!t)^2 - x^2} - 
\frac{1}{a_2^2 (T\!-\!t)^2 - x^2} \right] \; .
\end{aligned}
\end{equation}
Noting $\hat a(x,t) \equiv x/(T-t)$ the drift velocity of an agent within
region (1) in the $\sigma \to 0$ limit ($\hat a(x,t) \in [a_2,a_0])$, $t_d
= (T-t)$ the drift time and $t_\sigma = x^2/\sigma^2$ the diffusion
times, Eq.~(\ref{eq:phi-small-sigma}) applies under the condition
that:
\begin{align}
\frac{t_d}{t_\sigma} = \frac{\sigma^2 (T-t)}{ x^2} &\ll \left( 1 -
  \frac{a^2_0}{\hat a^2(x,t)} \right)^2  \label{cond:small-sigma-1a}\\ 
\frac{t_d}{t_\sigma}  = \frac{\sigma^2 (T-t)}{ x^2} &\ll \left( 1 -
  \frac{a^2_2}{\hat a^2(x,t)} \right)^2  \label{cond:small-sigma-1b}\\ 
\sigma ^2 &\ll \frac{(T - \bar t)}{2}  \left(\hat a^2(x,t)  - a^2_2
\right) \; . \label{cond:small-sigma-1c}
\end{align}

The two first conditions express that if generally speaking
Eq.~(\ref{eq:phi-small-sigma}) requires that the time of drift is much
shorter than the diffusion time, the requirement becomes more and more
stringent as $(x,t)$ get closer from the boundaries of region (1)
where $\hat a(x,t) \to a_0$ or $\hat a(x,t) \to a_2$.  The last
condition signal that Eq.~(\ref{eq:phi-small-sigma}) is valid only if
$T$ differs significantly from $\bar t$.

The calculation of $\phi(x,t)$ in region (3) in the small $\sigma$
limit proceeds essentially along the same lines.

\subsection{Uniform approximations}

The conditions (\ref{cond:small-sigma-1a})-(\ref{cond:small-sigma-1b})
express that the transition between regions (0) and (1) (i.e. $x
\simeq -a_0 (T\!-\!t)$) as well as the transition  between regions (1) and (2) 
(i.e. $x \simeq -a_2 (T\!-\!t)$) need to be treated a bit more carefully and
require the use of uniform approximations.

For $(\alpha,\beta,\gamma) \bar t \ll \sigma^2$, one way to derive
these uniform approximation is just to select the dominating
contribution of Eq.~(\ref{eq:phi-full-c}).  Indeed, deep in region (0)
(respectively deep in region(2)), one can check that the third term
(respectively the first term) of Eq.~(\ref{eq:phi-full-c}) in which
$\operatorname {erfc}$ is replaced by its asymptotic value 2 recovers
exacly Eq.~(\ref{eq:phi-col}).  Near $x = -a_0(T\!-\!t)$ or $x =
-a_2(T\!-\!t)$ the uniform approximation amounts to keep the full
dependence of the $\operatorname {erfc}$.

For  instance near  $x = -a_2T$
\begin{equation} \label{eq:phi-uniform}
\phi(x,t)   \simeq \frac{e^{-\alpha (T-\bar t)/\sigma^2}}{2} 
  \exp \left( \frac{a_2^2(T \! - \! t) + 2a_2x}{2 \sigma^2} \right)
       \operatorname{erfc} \left(\frac{-x - a_2 (T \! - \! t)}{\sqrt{2 \sigma^2
         (T \! - \! t)}} \right) \; . \\
\end{equation} 
This expression will interpolate smoothly between Eq.~(\ref{eq:phi-col})
and Eq.~(\ref{eq:phi-small-sigma}). [This latter can be seen as  being obtained
 using the large $x$ asymptotic $\operatorname
{erfc}(x) \simeq \exp(-x^2)/\sqrt{\pi x^2}$ for the four terms of
Eq.~(\ref{eq:phi-full-c})].

 \subsection{drift velocity}

With the knowledge of $\phi(x,t)$, the drift velocities in the small
sigma regime can be obtained from the spatial derivative of
$u(x,t) = - \sigma^2 \log \phi(x,t)$ (cf Eqs.~\eqref{eq:drift} and
\eqref{eq:u-general}).  In leading $\sigma$ order
Eqs.~\eqref{eq:phi-col} and \eqref{eq:phi-small-sigma} yield 
\begin{equation} \label{eq:u_small_sigma}
 u(x,t) \simeq
\begin{cases}
\displaystyle  \alpha (T-\bar t)  - \frac{a_{0}^2}{2} (T \! - \! t) -
   a_{0}x & \mbox{ in Region (0) } \\
\displaystyle     c(T) + \frac{x^2}{2(T-t)}  & \mbox{ in Region (1) } \\
\displaystyle \alpha (T-\bar t)   - \frac{a_{2}^2}{2} (T \! - \! t) -    a_{2}x	   &
\mbox{ in Region (2) } \\ 
    \end{cases}  \; ,
\end{equation}
(the expression of $u(x,t)$ for region~(3) can be obtained in the same
way).  Taking the spatial derivative of these expressions yields the
velocities Eq.~\eqref{eq:a_small_sigma}.

Alternatively, one can compute the drift velocity $a(x,t) =
\sigma^2 \partial_x \phi(x,t)/ \phi(x,t)$ from the spatial derivative
$\partial_x \phi(x,t)$, which can be evaluated following exactly the
same steps as for $\phi(x,t)$.  This of course gives the same result.

\bibliography{Biblio}

\end{document}